\documentclass[11pt,a4paper]{article}
\usepackage{sint,amsfonts}
\usepackage{booktabs}
\usepackage{amsmath}
\usepackage{amssymb}
\numberwithin{equation}{section}
\usepackage{lineno,hyperref}
\usepackage{bbold}
\usepackage{wasysym}
\usepackage[flushleft]{threeparttable}
\usepackage{lscape}
\usepackage{rotating}
\usepackage{simplewick}
\usepackage{upgreek}
\usepackage[english]{babel}
\usepackage{cite} 
\usepackage{xcolor}
\usepackage{multirow}
\usepackage{caption}

\def\rms{{\rm s}}

\def\sl{\rms_\lambda}

\def\bsx{{\boldsymbol x}}
\def\bsxi{{\boldsymbol \xi}}

\def\proof{\noindent{\sl Proof:}\kern0.6em}

\def\frac#1#2{\hbox{$#1\over#2$}}
\def\dual{\mathstrut^*\kern-0.1em}

\def\lvec#1{\setbox0=\hbox{$#1$}
    \setbox1=\hbox{$\scriptstyle\leftarrow$}
    #1\kern-\wd0\smash{
    \raise\ht0\hbox{$\raise1pt\hbox{$\scriptstyle\leftarrow$}$}}
    \kern-\wd1\kern\wd0}
\def\rvec#1{\setbox0=\hbox{$#1$}
    \setbox1=\hbox{$\scriptstyle\rightarrow$}
    #1\kern-\wd0\smash{
    \raise\ht0\hbox{$\raise1pt\hbox{$\scriptstyle\rightarrow$}$}}
    \kern-\wd1\kern\wd0}
\def\slash#1{\setbox0=\hbox{$#1$}\setbox1=\hbox{$\kern1pt/$}
    #1\kern-\wd0\kern1pt/\kern-\wd1\kern\wd0}

\def\nab#1{{\nabla_{#1}}}
\def\nabstar#1{{\nabla\kern0.5pt\smash{\raise 4.5pt\hbox{$\ast$}}
               \kern-5.5pt_{#1}}}

\def\nabbar#1{{\overleftarrow{\nabla}_{#1}}}
\def\nabbarstar#1{{\overleftarrow{\nabla}\kern0.5pt\smash{\raise 4.5pt\hbox{$\ast$}}
               \kern-5.5pt_{#1}}}

\def\nabdbar#1{{\overleftrightarrow{\nabla}_{#1}}}
\def\nabdbarstar#1{{\overleftrightarrow{\nabla}\kern0.5pt\smash{\raise 4.5pt\hbox{$\ast$}}
               \kern-5.5pt_{#1}}}

\def\drvstar#1{{\partial\kern0.5pt\smash{\raise 4.5pt\hbox{$\ast$}}
               \kern-6.0pt_{#1}}}

\def\ldrvstar#1{{\lvec{\,\partial}\kern-0.5pt\smash{\raise 4.5pt\hbox{$\ast$}}
               \kern-5.0pt_{#1}}}

\def\MSbar{\overline{\rm MS\kern-0.5pt}\kern0.5pt}

\def\psibar{\overline{\psi}}

\def\zetabar{\bar{\zeta}}
\def\zetaprime{\zeta\kern1pt'}
\def\zetabarprime{\zetabar\kern1pt'}

\def\dirac#1{\gamma_{#1}}
\def\diracstar#1#2{
    \setbox0=\hbox{$\gamma$}\setbox1=\hbox{$\gamma_{#1}$}
    \gamma_{#1}\kern-\wd1\kern\wd0
    \smash{\raise4.5pt\hbox{$\scriptstyle#2$}}}

\def\Tr{{\rm Tr}}

\def\Ds{D_{\rm s}}
\def\DsdagDs{\Ds{\Ds}^{\kern-1pt\dagger}}

\def\avg#1{{\kern1.0pt\overline{\kern-1.0pt#1\kern-1.0pt}\kern1.0pt}}

\newcommand{\be}{\begin{equation}}
\newcommand{\ee}{\end{equation}}
\newcommand{\bea}{\begin{eqnarray}}
\newcommand{\eea}{\end{eqnarray}}

\newcommand{\msbar}{{\rm \overline{MS\kern-0.05em}\kern0.05em}}

\newcommand{\ba}{\begin{eqnarray}}
\newcommand{\ea}{\end{eqnarray}}

\renewcommand{\vec}[1]{\boldsymbol{#1}}

\newcommand{\corr}[1]{{\langle#1\rangle}}

\newcommand{\thA}{\theta_0^A}
\newcommand{\thB}{\theta_0^B}
\newcommand{\sss}{\scriptscriptstyle}
\newcommand{\NuAB}{{{\cal V}_{0,k}^{AB}}}

\usepackage{xcolor}

\begin{document}

\begin{titlepage}
\begin{flushright}
CERN-TH-2026-139
\end{flushright}

  \begin{center}

{\Large\bf The QCD energy-momentum tensor on the lattice: non-perturbative renormalization
  with $N_f=3$\\[0.5ex]}

\end{center}
\vskip 0.75 cm
\begin{center}

 {\large Matteo Bresciani$^{\scriptscriptstyle a}$,
  Mattia Dalla Brida$^{\scriptscriptstyle b,c}$,
  Leonardo Giusti$^{\scriptscriptstyle b,c,d}$,\\
 Mitsuaki Hirasawa$^{\scriptscriptstyle b,c}$,
 Michele Pepe$^{\scriptscriptstyle c}$,
 and Luca Virz\`i$^{\scriptscriptstyle b,c}$}

\vskip 1.25cm
$^{\scriptstyle a}$ School of Mathematics and Hamilton Mathematics Institute, Trinity College Dublin,\\ Dublin 2, Ireland\\
$^{\scriptstyle b}$ Department of Physics “Giuseppe Occhialini”, University of Milano-Bicocca,\\Piazza della Scienza 3, I-20126 Milano, Italy\\
$^{\scriptstyle c}$ INFN Milano-Bicocca,\\ Piazza della Scienza 3, I-20126 Milano, Italy\\
$^{\scriptstyle d}$ Theoretical Physics Department, CERN, 1211 Geneva 23, Switzerland\\

\vskip 1.5cm
{\bf Abstract}
\vskip 0.35ex
\end{center}

\noindent
We construct the traceless components of the energy-momentum tensor on the lattice for QCD with $N_f=3$
flavours, such that their correlation functions satisfy the appropriate Ward identities in the continuum limit. To carry out
this program, we define the theory on the lattice by the Wilson-plaquette and the $O(a)$-improved Wilson actions
for gluons and quarks respectively. The discretization of the space-time entails that (i) the irreducible
nonet representation of the SO($4$) group splits into a triplet and a sextet irreducible representations of
the hypercubic group, and (ii) for each multiplet non-perturbative determinations of the the gluonic and
fermionic renormalization constants are required.  The bare gluonic components of the energy-momentum tensor are
defined via the clover discretization of the field strength tensor, while the fermionic ones are discretized
by appropriate combinations of symmetric covariant derivatives. Either for the triplet or the sextet representations,
the two independent renormalization constants are then fixed non-perturbatively by imposing discretized
versions of continuum Ward identities for one-point correlation functions in the presence of shifted boundary
conditions and an imaginary chemical potential. The non-perturbative calculation is then carried out by Monte Carlo
simulations, and the resulting renormalization constants are determined with a final accuracy of a few percent
for values of the bare coupling constant squared in the range $0 \leq g_0^2\leq 0.96$.
\vfill

\eject

\end{titlepage}

\tableofcontents

\section{Introduction} 

The energy-momentum tensor $T_{\mu\nu}$ has a central role in Quantum
Chromodynamics (QCD) as it encodes the response of the theory to space-time
transformations and to external gravitational 
fields~\cite{Weinberg:1972lg}, becoming a source of primordial 
gravitational waves emitted by the plasma in the early Universe.
It provides the conserved currents associated to the
Poincaré invariance, governs the realization of scale transformations, and couples the theory to variations of the background metric.
As such, its matrix elements give direct access to a wide range of physically relevant observables characterizing strongly interacting
matter. 

At zero temperature, matrix elements of $T_{\mu\nu}$ determine the gravitational form factors of hadrons. The trace (anomaly)
plays a key role in understanding the origin of  the hadron masses, while other components give a detailed
description of the internal structure, including the spatial distributions of the pressure and of the shear forces,
and information about the so-called D-term, see Ref.~\cite{Polyakov:2018zvc} for a comprehensive review.
All these quantities probe intrinsically non-perturbative aspects of QCD, and require a precise non-perturbative definition
of $T_{\mu\nu}$ satisfying the appropriate Ward identities.

At finite temperature, the energy-momentum tensor is the link between microscopic dynamics and macroscopic properties
of the QCD plasma. Thermodynamic quantities such as the pressure, the energy density, the entropy density and the trace anomaly are obtained from
thermal expectation values of the various components of $T_{\mu\nu}$ \cite{Landau-hydro}. Beyond equilibrium thermodynamics,
correlation functions of
the energy-momentum tensor encode transport properties of the medium. Through Kubo relations, shear and bulk viscosities can be extracted
from the zero-frequency limit of the corresponding spectral functions~\cite{Kubo:1957mj,Kubo:1957wcy}. A precise non-perturbative definition
of $T_{\mu\nu}$ is therefore indispensable not only for the study of the Equation of State but also for establishing a theoretically
sound connection between Euclidean lattice correlators and real-time observables.

While in the continuum the definition and the conservation of the energy-momentum tensor follow directly from space-time symmetries,
its non-perturbative definition on the lattice presents significant challenges. The discretization of space-time breaks explicitly
the SO($4$) invariance down to the hypercubic group and, as a consequence, the irreducible
nonet representation of the traceless part of $T_{\mu\nu}$ splits into a triplet and a sextet representations of
the hypercubic group. In addition, each multiplet is neither exactly conserved nor satisfies
exact Ward identities at finite spacing, and its gluonic and
fermionic components mix under renormalization~\cite{Caracciolo:1989pt,Caracciolo:1989bu}. A non-perturbative definition of the
(renormalized) energy-momentum tensor must therefore rely on the explicit enforcement of
discretized versions of continuum Ward identities which become exact in the continuum limit\cite{Caracciolo:1989pt}. 

A particularly effective strategy to address these issues was proposed in Refs.~\cite{Giusti:2015daa,DallaBrida:2020gux}, where
the thermal quantum field theory is formulated in a moving reference frame by imposing shifted boundary
conditions~\cite{Giusti:2010bb,Giusti:2011kt,Giusti:2012yj} in the presence of an imaginary chemical
potential~\cite{DallaBrida:2020gux}. In this framework, exact Ward identities for one-point functions can be derived which, once enforced
on the lattice up to discretization effects which vanish in the continuum limit, define the renormalization conditions
for the energy-momentum tensor. This approach has proven to be especially efficient from the numerical view point in
the SU($3$) Yang--Mills theory, where a high precision was achieved in the
determination of the renormalization constants~\cite{Giusti:2015daa}.

The purpose of this work is to generalize the results in Ref.~\cite{Giusti:2015daa} to QCD by implementing the strategy
proposed in Ref.~\cite{DallaBrida:2020gux} for defining
non-perturbatively the QCD energy-momentum tensor on the lattice. To this aim, we consider the plaquette Wilson
and the $O(a)$-improved Wilson actions to discretize gluons and fermions, respectively. The bare gluonic components
of the energy-momentum tensor are defined via the clover discretization of the field strength tensor, while the
fermionic ones are discretized by appropriate combinations of symmetric covariant derivatives.  We focus on the
traceless components of $T_{\mu\nu}$ and determine the renormalization constants of both the triplet and the sextet
gluonic and fermionic contributions with percent precision.

For completeness, we notice that alternative non-perturbative approaches based on the gradient flow~\cite{Luscher:2010iy} have also been
attempted~\cite{Suzuki:2013gza,DelDebbio:2013zaa,Makino:2014taa,Capponi:2015ahp,Taniguchi:2016ofw,Harlander:2018zpi,PavanPavan:2025hzr}.
In that framework, flowed composite fields are used to define non-perturbatively the energy-momentum tensor by extrapolating
the lattice results first to the continuum limit and then to zero flow-time.

\section{Renormalization strategy}
\label{sec:theory}
The renormalization procedure starts by identifying continuum
Ward identities associated with space-time translations which,
once discretized, can be enforced efficiently by numerical simulations.
Their discrete counterparts are then imposed on the lattice so as to
fix the renormalization constants of the various components
of the energy-momentum tensor. While the derivation of these
identities can be found in Ref.~\cite{DallaBrida:2020gux},
here we present a self-contained discussion focused on the theoretical
setup underlying the strategy\footnote{We refer to Ref.~\cite{DallaBrida:2020gux} for all quantities
not explicitly defined here.}. 

\subsection{Continuum Ward identities}
The Ward identities we are interested in can be derived by formulating
the partition function of a thermal system in a moving
reference frame through shifted boundary
conditions~\cite{Giusti:2010bb,Giusti:2011kt,Giusti:2012yj} combined 
with the presence of an imaginary chemical
potential~\cite{DallaBrida:2020gux}. For the gauge field, shifted 
boundary conditions along the compact direction of size $L_0$ read
\begin{equation}
  A_\mu(x_0 + L_0, \bsx) = A_\mu(x_0, \bsx - L_0\,\bsxi)\, ,
\end{equation}
while for the fermion fields the boundary conditions are 
\begin{equation}\label{eq:shiftedBC}
  \psi(x_0 + L_0, \bsx) = -e^{i\theta_0}\,\psi(x_0, \bsx - L_0\,\bsxi)\,,
  \quad
\psibar(x_0+L_0,\bsx)  =  -e^{-i \theta_0}\, \psibar(x_0,\bsx - L_0\bsxi)\, ,
\end{equation}
where $\bsxi$ is the shift vector that specifies the (imaginary) velocity of the moving frame,
and $\theta_0$ is a fermionic twist phase playing the role of an imaginary chemical
potential, $\mu_I= -\theta_0/L_0$. All the fields satisfy periodic boundary 
conditions along the three spatial directions, each of length $L$. The resulting partition function $Z (L_0,\bsxi,\theta_0)$ and
the associated free-energy density $f (L_0,\bsxi,\theta_0)$ are defined as
\begin{equation}\label{eq:shiftedsys}
  Z (L_0,\bsxi,\theta_0) = \int {\cal D} U\, {\cal D} \psi\, {\cal D} \psibar\, e^{-S}\,,
  \qquad
  f (L_0,\bsxi,\theta_0) = -\dfrac{1}{L_0 L^3} \ln Z (L_0,\bsxi,\theta_0)\, , 
\end{equation}
where $S$ is the QCD action. The interplay between the baryonic symmetry and 
the $\mathbb{Z}(3)$ center symmetry of the gauge sector, combined 
with the fact that the free-energy density is even
under a sign flip of $\theta_0$ (charge conjugation), restricts the twist
phase $\theta_0$ to the range $[0,\pi/3]$~\cite{Roberge:1986mm}.  

The energy-momentum tensor $T_{\mu\nu}$ is a dimension-4 gauge invariant symmetric two-index tensor,
see Ref.~\cite{DallaBrida:2020gux} for its definition, which satisfies 
the identity
\begin{equation}
  \langle T_{0k} \rangle_{\boldsymbol{\xi},\theta_0} =
  - \dfrac{\partial}{\partial \xi_k} f (L_0,\bsxi,\theta_0)\,,
  \label{eq:WI_shift_cont} 
\end{equation}
relating the thermal expectation value of its space-time components
to the derivative of the free-energy density with respect to the
shift~\cite{Giusti:2010bb,Giusti:2011kt,Giusti:2012yj}. Additional Ward
identities of interest for this paper link diagonal and off-diagonal matrix
elements as given by the following equations
\ba
\langle T_{0k} \rangle_{\sss\vec\xi,\theta_0} & = & \dfrac{\xi_k}{1-\xi_k^2} 
\Big\{\langle T_{00} \rangle_{\sss\vec\xi,\theta_0}  
- \langle T_{kk} \rangle_{\sss\vec \xi,\theta_0}\,\Big\}\, , \label{eq:WIodd}\\[0.25cm]
\langle T_{0k} \rangle_{\sss\vec\xi,\theta_0} & = & \xi_k 
\Big\{\langle T_{00} \rangle_{\sss\vec\xi,\theta_0}  
- \langle T_{jj} \rangle_{\sss\vec \xi,\theta_0}\,\Big\}\qquad (j\neq k, \xi_j=0 )\; .
\label{eq:WIodd2}
\ea
Finally, conservation of the baryon number implies the following identity
\begin{equation}
  \langle V_0 \rangle_{\sss\boldsymbol{\xi},\theta_0}  =
  - i L_0 \dfrac{\partial}{\partial \theta_0} f (L_0,\bsxi,\theta_0)\, ,
  \label{eq:WI_theta_cont}
\end{equation}
which relates the expectation value of the temporal component of the  flavour-singlet vector current
to the derivative of the free energy density with respect to the twist angle. Altogether the
above relations, once discretized, provide a sufficient set of identities to fix the renormalization constants
of the traceless components of the energy-momentum tensor on the lattice.  Notice that,
thanks to the presence of the shift, only one-point functions of $T_{\mu\nu}$ are involved. This
makes the numerical computation on the lattice significantly cheaper with respect to the
calculation of multi-point correlation functions which would be needed without the shift. 

\subsection{Lattice energy-momentum tensor and renormalization conditions}
We consider QCD on a lattice with spacing $a$ by discretizing gluons and fermions with 
the plaquette and the $O(a)$-improved Wilson actions, see Appendix~\ref{App:action} and
Ref.~\cite{DallaBrida:2020gux} for all conventions and definitions.
Bosonic and fermionic fields satisfy shifted boundary conditions in complete analogy with
the continuum formulation, the only difference being the replacement of $A_\mu$ with the link field $U_\mu$.
The components of $L_0\, \bsxi$ are now integers in units of $a$.

As a consequence of the breaking of the SO($4$) symmetry at finite lattice spacing,
the traceless part of a symmetric two-index tensor decomposes into the sum of two
irreducible representations of the hypercubic group: a sextet and a triplet.
The off-diagonal components belong to the sextet,
while the triplet consists of traceless linear combinations of the diagonal components.
This decomposition applies independently to the gluonic and to the fermionic components, i.e.  
\begin{equation}
  T^G_{\mu\nu} \rightarrow T^{G,\{6\}}_{\mu\nu} + T^{G,\{3\}}_{\mu\nu}\, ,
  \qquad 
  T^F_{\mu\nu} \rightarrow T^{F,\{6\}}_{\mu\nu} + T^{F,\{3\}}_{\mu\nu} \, .
\end{equation}
The bare gluonic fields are given by
\ba
  {T}_{\mu\nu}^{G,\{6\}} & = & \big(1-\delta_{\mu\nu})
  \dfrac{1}{g_0^2}\Big\{{F}^a_{\mu\alpha} {F}^a_{\nu\alpha} \Big\}\; ,\label{eq:TG6}\\[0.25cm]
  {T}_{\mu\nu}^{G,\{3\}} & = & \dfrac{1}{g_0^2}\Big\{{F}^a_{\mu\alpha} {F}^a_{\mu\alpha} - {F}^a_{\nu\alpha} {F}^a_{\nu\alpha}\Big\}\; ,
  \label{eq:TG3}
\ea
where $g_0$ is the bare coupling, ${F}_{\mu\nu} = {F}_{\mu\nu}^a T^a$ with ${F}_{\mu\nu}^a=2\,\Tr\{\widehat{F}_{\mu\nu} T^a\}$, and
$T^a$ are the Hermitian generators of the SU($3$) gauge group normalized as $\Tr \{ T_a T_b\}= \delta_{ab}/2$. The bare fermionic operators are 
\ba
{T}_{\mu\nu}^{F,\{6\}} & = &  (1-\delta_{\mu\nu})
  \dfrac{1}{8}\left\{\psibar\dirac\mu\big[\nabdbarstar\nu+\nabdbar\nu\big]\psi +
  \psibar\dirac\nu\big[\nabdbarstar\mu+\nabdbar\mu\big]\psi \right\}\; ,\label{eq:TF6}\\[0.25cm]
{T}_{\mu\nu}^{F,\{3\}} & = & \dfrac{1}{4}\Big\{\psibar\dirac\mu\left[\nabdbarstar\mu+\nabdbar\mu\right]\psi
- \psibar\dirac\nu\left[\nabdbarstar\nu+\nabdbar\nu\right]\psi\Big\}\; , \label{eq:TF3}
\ea
where the lattice covariant derivatives, $\nabdbar\mu$ and $\nabdbarstar\mu$, and $\widehat{F}_{\mu\nu}$
are defined in Appendix~\ref{App:action}.

The standard analysis for the mixing of composite fields shows 
that, for each representation, the renormalized fields are linear combinations
of the corresponding bare gluonic and
fermionic components~\cite{Caracciolo:1989pt,Caracciolo:1989bu}
\begin{equation}\label{eq:renormEMT}
  {T}_{\mu\nu}^{R,\{i\}} =  Z^{\{i\}}_{G}(g_0^2)\, {T}_{\mu\nu}^{G,\{i\}} +
  Z^{\{i\}}_{F}(g_0^2)\, {T}_{\mu\nu}^{F,\{i\}}\; ,
\end{equation}
with two pairs of independent renormalization constants, $Z^{\{i\}}_{G}(g_0^2)$ and
$Z^{\{i\}}_{F}(g_0^2)$, one for the sextet ($i=6$) and one for the
triplet ($i=3$). These renormalization constants have to be fixed non-perturbatively by imposing a sufficient number of
continuum Ward identities discretized on the lattice.
The renormalization conditions that we adopt in this paper are a discretized version of
Eqs.~(\ref{eq:WI_shift_cont})-(\ref{eq:WIodd2}), namely 
\begin{equation}\label{eq:rencond0}
  \langle {T}_{0k}^{R,\{6\}} \rangle_{\scriptscriptstyle \vec\xi,\theta_0} =  - \dfrac{\Delta}{\Delta \xi_k} f (L_0,\bsxi,\theta_0) \, ,
\end{equation}
where $\Delta / \Delta \xi_k$ denotes the lattice discretization of the derivative, and
\ba
\langle T_{0k}^{R,\{6\}} \rangle_{\sss\vec\xi,\theta_0} & = & \dfrac{\xi_k}{1-\xi_k^2} \;
\langle T_{0k}^{R,\{3\}} \rangle_{\sss\vec\xi,\theta_0}  
\; ,\label{eq:WIoddLAT}\\[0.25cm]
\langle T_{0k}^{R,\{6\}} \rangle_{\sss\vec\xi,\theta_0} & = & \xi_k \;
\langle T_{0j}^{R,\{3\}} \rangle_{\sss\vec\xi,\theta_0}\, ,  
\qquad (j\neq k, \xi_j=0 )\, .
\label{eq:WIodd2LAT}
\ea
We notice that Eq.~(\ref{eq:WIodd2LAT}) is especially useful when Eq.~(\ref{eq:WIoddLAT}) becomes singular, as for the case $\xi_k =1$. The
above equations involve only one-point functions of ${T}_{\mu\nu}^{R,\{i\}}$. This represents a crucial
numerical advantage, yielding significantly
higher statistical precision compared to Ward identities involving multi-point correlation functions.
In order to
determine $Z^{\{i\}}_{G}(g_0^2)$ and $Z^{\{i\}}_{F}(g_0^2)$, one needs two independent equations for each hypercubic representation.
Since $\theta_0$  couples directly to fermions while it affects gluons only indirectly through fermion loops, 
the use of different values of the fermionic twist phase achieves the goal of disentangling the gluonic and fermionic contributions.
Labeling the two twist phases by $\theta_0^A$ and
$\theta_0^B$, we obtain the following system of two equations,
\begin{equation}\label{eq:rencond}
\begin{cases}
  {\cal Z}^{\{6\}}_{G}\, \langle {T}_{0k}^{G,\{6\}} \rangle_{\scriptscriptstyle \vec\xi,\theta_0^A}
  + {\cal Z}^{\{6\}}_{F}\, \langle {T}_{0k}^{F,\{6\}} \rangle_{\scriptscriptstyle \vec\xi,\theta_0^A} =
  - \dfrac{\Delta}{\Delta \xi_k} f (L_0,\bsxi,\theta_0^A)
  \\ \\
  {\cal Z}^{\{6\}}_{G}\, \langle {T}_{0k}^{G,\{6\}} \rangle_{\scriptscriptstyle \vec\xi,\theta_0^B}
  + {\cal Z}^{\{6\}}_{F}\, \langle {T}_{0k}^{F,\{6\}} \rangle_{\scriptscriptstyle \vec\xi,\theta_0^B} =
  - \dfrac{\Delta}{\Delta \xi_k} f (L_0,\bsxi,\theta_0^B)\,,
\end{cases}
\end{equation}
whose solutions ${\cal Z}^{\{6\}}_{G}$ and ${\cal Z}^{\{6\}}_{F}$
depend\footnote{In general we omit to indicate the dependences on the 
various parameters for the clarity of the notation except when showing
(some of) them helps the presentation.} on the bare coupling
constant~\cite{Caracciolo:1989pt,Caracciolo:1989bu}, while their (scheme
and scale) dependence on $L_0$, on $\bsxi$, and on $\theta_0$ is only via
irrelevant discretization effects\footnote{In this sense it is often said
that these renormalization constants are scheme and scale independent.}
which, notably, start at $O(a^2)$~\cite{DallaBrida:2020gux}.
The two renormalization constants for the triplet representation, 
${\cal Z}^{\{3\}}_{G}$ and ${\cal Z}^{\{3\}}_{F}$, can be derived similarly  
using Eqs.~(\ref{eq:WIoddLAT}) or (\ref{eq:WIodd2LAT}).
At fixed $g_0^2$, the leading-order $L_0$-dependent lattice artifacts can be parameterized as follows,
\begin{equation}
\displaystyle
    {\cal Z}_X^{\{i\}}(g_0^2,a/L_0,\bsxi,\theta_0) = Z_X^{\{i\}}(g_0^2) + \dfrac{a}{L_0}\, \Big\{   
    z_{X,1}^{\{i\}}\, \Big(\dfrac{a}{L_0}\Big) 
  + z_{X,2}^{\{i\}}\, (a\Lambda_{\rm QCD}) + O(a^2)\Big\}\,,
    \label{eq:aTto0_theory}
\end{equation}
where $i=3,6$, $X=F,G$ stands for the fermionic and gluonic cases, and the
dependence of the coefficients $z_{X,1}^{\{i\}}$ and $z_{X,2}^{\{i\}}$
on the various parameters is omitted.
A possible choice for the renormalization scheme is to define the 
renormalization constants
at fixed $L_0$ as they stand. The residual dependence on $L_0$, $\bsxi$ and $\theta_0$ will
eventually contribute to the $O(a^2)$ effects of a renormalized correlation
function of the energy-momentum tensor. In practice, however, these discretization effects
turn out to be quite small, see below. We therefore choose to define
the renormalization constants of the traceless components of the energy-momentum
tensor as\footnote{Notice that the linear term in $a/L_0$ in Eq.~(\ref{eq:aTto0_theory})
is suppressed by the factor
$a\Lambda_{\rm QCD}$, where $\Lambda_{\rm QCD}=344(8)$~MeV in the 
$\overline{\rm MS}$ scheme~\cite{DallaBrida:2026kuo}, which is always (very) small at
the values of $\beta$ considered in this study.}
\begin{equation}
Z^{\{i\}}_{G}(g_0^2) =\!\! \lim_{1/L_0\rightarrow 0}\!\! {\cal Z}^{\{i\}}_{G}(g_0^2,a/L_0,\bsxi,\theta_0)\, , \quad
Z^{\{i\}}_{F}(g_0^2) =\!\!\lim_{1/L_0\rightarrow 0}\!\! {\cal Z}^{\{i\}}_{F}(g_0^2,a/L_0,\bsxi,\theta_0)\, ,
\label{eq:liml02inf}
\end{equation}
where the limit is taken at fixed $g_0$, $\bsxi$ and $\theta_0$. This removes the
$L_0$-dependent discretization effects as well as the dependence on
$\bsxi$ and $\theta_0$. $L_0$-independent discretization effects
proportional to $(a\Lambda_{\rm QCD})^p$, with $p \geq 2$, remain as part of the definition of
the renormalization constants. Again, they will eventually contribute
to the $O(a^2)$ effects of a renormalized correlation function of the
energy-momentum tensor, and will be removed when the continuum
limit of the latter is taken.

Notice that, once Eqs.~(\ref{eq:rencond}), the analogous ones for the triplet, and (\ref{eq:liml02inf})
have been enforced, it is guaranteed that all correlators of the renormalized energy-momentum tensor defined in
Eq.~(\ref{eq:renormEMT}) satisfy the proper Ward identities in the continuum limit, i.e. the
non-perturbatively defined traceless part of the energy-momentum tensor is conserved in the continuum limit.

An additional simplification arises from the exact baryon number conservation on the lattice with Wilson fermions.
The
corresponding Ward
identity, Eq.~(\ref{eq:WI_theta_cont}), is satisfied by the exactly conserved lattice flavour-singlet
vector current
\begin{equation}\label{eq:VectorCurrent}
        V_0 (x) = \dfrac{1}{2} \left\{\psibar(x+ a\, \hat{0}) U_0^\dag(x)(\gamma_0+1)\psi(x) +
                                          \psibar(x) U_0(x)(\gamma_0-1)\psi(x+ a\, \hat{0})\right\} \, ,
\end{equation}
a fact which allows one to relate the free-energy derivatives evaluated at two different twist phases $\theta_0^A$ and $\theta_0^B$
as follows
\begin{equation}\label{eq:dfB}
\dfrac{\Delta}{\Delta \xi_k} f (L_0,\bsxi,\theta_0^B) = \dfrac{\Delta}{\Delta \xi_k} f (L_0,\bsxi,\theta_0^A) - {\cal V}_{0,k}^{AB}\, ,
\end{equation}
where
\begin{equation}\label{eq:V_{0,k}^{AB}}
  {\cal V}_{0,k}^{AB} = -\dfrac{i}{L_0} \int_{\theta_0^A}^{\theta_0^B} d\theta_0 \, \dfrac{\Delta}{\Delta \xi_k}
  \langle V_0\rangle_{\scriptscriptstyle \vec\xi,\theta_0}\, .
\end{equation}
This equation plays an important practical role in the numerical determination of the renormalization constants, as it reduces the need
for direct (computationally expensive) evaluations of free-energy derivatives at multiple twist angles. Finally, as in the continuum theory, the
physically relevant range of the twist phase remains $[0,\pi/3]$ as a consequence of the combined symmetry properties of the fermionic
sector and of the $\mathbb{Z} (3)$ center symmetry of the gauge theory.

\section{The numerical study}
\label{sec:numsimulations}
The computation of the renormalization constants using 
Eqs.~\eqref{eq:rencond}
(and the analogous ones for the triplet representation) requires the lattice
determination of the matrix elements of the bare energy-momentum tensor and
of the discrete derivatives of the free-energy density with respect to the shift.
We have therefore simulated lattice QCD with the Wilson plaquette gauge
action and $N_f=3$ flavours of massless, $O(a)$-improved Wilson fermions. In order
to re-use previously generated gauge-field ensembles and results, the temporal
sizes $L_0/a=4,6,8$, the shift $\bsxi=(1,0,0)$, and the values of the inverse bare
coupling squared $\beta=6/g_0^2$ have been chosen as in
Refs.~\cite{DallaBrida:2021ddx,Bresciani:2025vxw,Bresciani:2025mcu},
where the values of the improvement coefficient $c_{\rm sw}$ and of the critical
quark mass are given as well. To those ensembles, we have added lattices at the bare
parameters $L_0/a=4,6$ and $\beta=15$.

As discussed in Section~\ref{sec:theory}, Eqs.~\eqref{eq:rencond} are
evaluated at two different values of the fermionic
twist angle. While the statistical error on
$\langle {T}_{\mu\nu}^{G,\{i\}} \rangle_{\scriptscriptstyle \vec\xi,\theta_0}$ and
$\langle {T}_{\mu\nu}^{F,\{i\}} \rangle_{\scriptscriptstyle \vec\xi,\theta_0}$ is mostly
independent on the twist angle, the central value of their difference is enhanced
when $(\theta_0^B - \theta_0^A)$ is large, thus improving the numerical precision of the
solution of the linear system with respect to the renormalization constants.
Therefore, in order to maximize the effect of $\theta_0$, we have chosen two 
well-separated phases within the allowed range, namely $\theta_0^A=0$ and 
$\theta_0^B=3\pi/10$. 

Monte Carlo simulations have been performed with the algorithms and the code described
in Refs.~\cite{DallaBrida:2021ddx,Bresciani:2025mcu}. High statistics is used to ensure
an accurate and robust determination of both the expectation values of the energy-momentum
tensor matrix elements and the finite difference derivative of the free-energy density.
Errors have been estimated and propagated from primary observables to the renormalization
constants using the Gamma-method~\cite{Wolff:2003sm} in the implementation of 
Refs.~\cite{Ramos:2018vgu,Joswig:2022qfe}.

\subsection{Finite-volume effects}
The basic equations~\eqref{eq:WI_shift_cont}--\eqref{eq:WI_theta_cont}
are valid in finite volume. To optimize the numerical cost, however,
we have computed the various quantities entering
these identities at different (very large) volumes. Since thermal QCD enjoys
exponential suppression of finite-volume effects, with the 
exponent being proportional to $L/L_0$~\cite{Giusti:2012yj,Laine:2009dh},
we have chosen the length of the spatial directions, $L$, so 
that $18\lesssim L/L_0\lesssim 70$.
Given these aspect ratios, we
expect finite-volume effects to be completely negligible with respect to the
statistical uncertainty of the Monte Carlo data. Indeed, the comparison of
results for the matrix elements of the bare energy-momentum tensor
determined on lattices with spatial sizes $L/a=144$ and $L/a=288$
shows no statistically significant difference~\cite{Bresciani:2025mcu,Bresciani:2025vxw},
thus confirming the theoretical expectation.

\subsection{Restricting to the trivial topological sector}
In QCD at high temperature, the topological charge distribution is strongly
peaked at the trivial sector. The instanton semi-classical analysis predicts the 
topological susceptibility of QCD with three light quark flavours to be
suppressed with the third power of the quark mass and approximatively
with the eighth power of the inverse temperature. The 
analogous prediction for the pure Yang-Mills case has been explicitly 
verified non-perturbatively~\cite{Giusti:2018cmp}. Similarly, lattice QCD
results tend to confirm the semiclassical 
prediction~\cite{Bonati:2015vqz,Borsanyi:2016ksw}, even though simulations 
with dynamical quarks are more challenging. The thermal ensembles 
considered in this work have physical temperatures larger than $5$~GeV, thus
we can safely restrict our measurements to the trivial sector as any 
effect on the observables from non-vanishing topological sectors is 
completely negligible with respect to the statistical
fluctuations~\cite{DallaBrida:2021ddx,Bresciani:2025mcu}.

\subsection{Computation of $\langle {T}_{\mu\nu}^{G,\{i\}} \rangle_{\scriptscriptstyle \vec\xi,\theta_0^A}$
and $\langle {T}_{\mu\nu}^{F,\{i\}} \rangle_{\scriptscriptstyle \vec\xi,\theta_0^A}$}
\label{ssec:T0k_thA}
The thermal expectation values of the bare gluonic and fermionic traceless 
components of the energy-momentum tensor at $\theta_0^A=0$ have been 
measured on lattices with spatial size of $L/a=288$~\cite{DallaBrida:2021ddx}
except for those at $\beta=15$ which have $L/a=144$.

For the triplet $T_{\mu\nu}^{G,\{3\}}$, 
see Eq.~\eqref{eq:TG3}, we have profited from the degeneracy of the 
component $F_{0\alpha}^aF_{0\alpha}^a$ with $F_{1\alpha}^a F_{1\alpha}^a$, 
and of $F_{2\alpha}^aF_{2\alpha}^a$ with $F_{3\alpha}^aF_{3\alpha}^a$, 
occurring at our choice of the shift parameter, to average them so as to 
reduce the statistical fluctuations. The same degeneracy holds for the 
fermion components of $T_{\mu\nu}^{F,\{3\}}$, see Eq.~\eqref{eq:TF3}.

The numerical results are reported in Table~\ref{tab:tensor_thA} of 
Appendix~\ref{App:tables}, where the details on the collected statistics
and on the attained precision are provided too.

\subsection{Computation of
  $\langle {T}_{\mu\nu}^{G,\{i\}} \rangle_{\scriptscriptstyle \vec\xi,\theta_0^B}$ and
  $\langle {T}_{\mu\nu}^{F,\{i\}} \rangle_{\scriptscriptstyle \vec\xi,\theta_0^B}$}
The same observables as in Subsection~\ref{ssec:T0k_thA} have been 
computed at the same values of $L_0/a$ and $\beta$ but at the second twist 
phase $\theta_0^B=3\pi/10$ through dedicated simulations of lattices with
spatial size $L/a=144$. The generation of these ensembles has been carried 
out along the lines of Appendix~E of Ref.~\cite{DallaBrida:2021ddx}. The 
results are given in Table~\ref{tab:tensor_thB} of 
Appendix~\ref{App:tables}.

\subsection{Computation of $\frac{\Delta}{\Delta\xi_k}f(L_0,\bsxi,\theta_0^A)$}
The right-hand side of the renormalization condition in 
Eq.~\eqref{eq:rencond0} requires the numerical evaluation of the 
derivative of the free-energy density with respect to the shift at fixed values of 
$L_0/a$ and $\beta$. We have discretized this derivative using the 
two-point symmetric approximation
\begin{equation}
    \dfrac{\Delta }{\Delta\xi_k} f (L_0, \bsxi,\theta_0)= 
    \dfrac{L_0}{4a}\left[ f (L_0, \bsxi+\frac{2a}{L_0}\hat{k},\theta_0) 
    - f (L_0, \bsxi-\frac{2a}{L_0}\hat{k},\theta_0)\right]\,,
    \label{eq:f_diff_discrete_shift}
\end{equation}
which is accurate up to $O(a^2)$ effects. For the $\theta_0^A=0$ 
case, this quantity was already computed with high precision in
Refs.~\cite{Bresciani:2025vxw,Bresciani:2025mcu} in the context of the 
determination of the QCD Equation of State on lattices with spatial extent $L/a=144$.
At the bare parameters
$L_0/a=4,6$ and $\beta=15$, we have computed it
through dedicated lattice simulations
following the same strategy.
All numerical results are reported in
Table~\ref{tab:dfdv} of Appendix~\ref{App:tables}.

\subsection{Computation of ${\cal V}_{0,k}^{AB}$}\label{ssec:NuAB}
The derivative of the free-energy density at the twist angle 
$\theta_0^B=3\pi/10$ has been obtained by using the identity 
given in Eq.~(\ref{eq:dfB}). At fixed bare parameters, we have evaluated
numerically the integral ${\cal V}_{0,k}^{AB}$ defined in 
Eq.~(\ref{eq:V_{0,k}^{AB}}) using Gaussian quadratures. Since 
$f(L_0, \bsxi,\theta_0)$ is an even function of $\theta_0$, the thermal
expectation value $\langle V_0 \rangle_{\scriptscriptstyle \bsxi,\theta_0}$ 
is odd and, thus, a polynomial expansion includes only odd powers
of $\theta_0$. This suggests the change of variable $z=\theta_0^2$ in order
to attain the same numerical accuracy using only half of the points,
\begin{equation}\label{V_{0,k}^{AB}num}
{\cal V}_{0,k}^{AB} = -\dfrac{i}{L_0} \int_{z^A}^{z^B} dz \, \dfrac{1}{2\sqrt{z}}\dfrac{\Delta }{\Delta \xi_k}
  \langle V_0 \rangle_{\scriptscriptstyle \vec\xi,\sqrt{z}},
\end{equation}
where $z^A=(\theta_0^A)^2$ and $z^B=(\theta_0^B)^2$. For the values
considered in this work, the integrand is regular in $z$: in particular, 
$\theta_0^A=0$ corresponds to a point where 
$\langle V_0\rangle_{\scriptscriptstyle \vec\xi,\sqrt{z}}$ vanishes 
linearly in $\sqrt{z}$ so that the ratio remains finite.
The integral is evaluated numerically using a 3-point Gauss-Legendre 
quadrature in the interval $[z^A,z^B]$. Given the smoothness of the 
integrand function, this simple integration scheme is enough for the 
systematic effects from the numerical quadrature to be completely negligible
with respect to the statistical accuracy of the final results, see 
Appendix~\ref{app:NuAB_int} for a detailed discussion. The finite difference
appearing in Eq.~\eqref{V_{0,k}^{AB}num} has been discretized analogously
to the one for the free-energy in Eq.~\eqref{eq:f_diff_discrete_shift}. 
Thus, at given values of $L_0/a$ and $\beta$, the numerical evaluation of
the integral requires the determination of the expectation value
$\langle V_0 \rangle_{\scriptscriptstyle \vec\xi,\theta_0}$ of the
conserved vector current at the three twist phases $\theta_0=\sqrt{z}$ 
given by the nodes of the Gauss-Legendre quadrature, each at the two 
shift values $\bsxi=(1\pm\frac{2a}{L_0},0,0)$ for the discrete
derivative in the shift, for a total of 6 independent simulations.
These expectation values have been computed on lattices with spatial size
$L/a=144$. The numerical results for
$\Delta \langle V_0 \rangle_{\scriptscriptstyle \vec\xi,\theta_0}/\Delta\xi_k$,
and the corresponding estimates for ${\cal V}_{0,k}^{AB}$, are given in
Table~\ref{tab:dfdv} in Appendix~\ref{App:tables}, where the details
about the collected statistics and the attained precision are reported as well.

\section{Computation of the renormalization constants}\label{sec:compt_zs}
The renormalization constants of 
the sextet components of the energy-momentum tensor follow from the 
solution of the system in
Eqs.~\eqref{eq:rencond}, while for the triplet they are the solution of an
analogous linear system, obtained 
combining Eqs.~\eqref{eq:WIodd2LAT} and~\eqref{eq:rencond}. At our choice
of shift, $\bsxi=(1,0,0)$, the explicit expressions are  
\begin{equation}\label{eq:ZG6}
  {\cal Z}^{\{6\}}_{G} = \dfrac{
    \langle {T}_{01}^{F,\{6\}} \rangle_{\scriptscriptstyle \vec\xi,\theta_0^A}
    \left[ \dfrac{\Delta}{\Delta \xi_1} f (L_0,\bsxi,\theta_0^B)\right]
    -
    \langle {T}_{01}^{F,\{6\}} \rangle_{\scriptscriptstyle \vec\xi,\theta_0^B}
    \left[ \dfrac{\Delta}{\Delta \xi_1} f (L_0,\bsxi,\theta_0^A)\right]
  }{
    \langle {T}_{01}^{G,\{6\}} \rangle_{\scriptscriptstyle \vec\xi,\theta_0^A}\;
    \langle {T}_{01}^{F,\{6\}} \rangle_{\scriptscriptstyle \vec\xi,\theta_0^B}
    -
    \langle {T}_{01}^{G,\{6\}} \rangle_{\scriptscriptstyle \vec\xi,\theta_0^B}\;
    \langle {T}_{01}^{F,\{6\}} \rangle_{\scriptscriptstyle \vec\xi,\theta_0^A}
  }
\end{equation}
and
\begin{equation}\label{eq:ZF6}
  {\cal Z}^{\{6\}}_{F} = \dfrac{
    \langle {T}_{01}^{G,\{6\}} \rangle_{\scriptscriptstyle \vec\xi,\theta_0^B}
    \left[ \dfrac{\Delta}{\Delta \xi_1} f (L_0,\bsxi,\theta_0^A)\right]
    -
    \langle {T}_{01}^{G,\{6\}} \rangle_{\scriptscriptstyle \vec\xi,\theta_0^A}
    \left[ \dfrac{\Delta}{\Delta \xi_1} f (L_0,\bsxi,\theta_0^B)\right]
  }{
    \langle {T}_{01}^{G,\{6\}} \rangle_{\scriptscriptstyle \vec\xi,\theta_0^A}\;
    \langle {T}_{01}^{F,\{6\}} \rangle_{\scriptscriptstyle \vec\xi,\theta_0^B}
    -
    \langle {T}_{01}^{G,\{6\}} \rangle_{\scriptscriptstyle \vec\xi,\theta_0^B}\;
    \langle {T}_{01}^{F,\{6\}} \rangle_{\scriptscriptstyle \vec\xi,\theta_0^A}
  }
\end{equation}
for the sextet case, and
\begin{equation}\label{eq:ZG3}
  {\cal Z}^{\{3\}}_{G} = \dfrac{
    \langle {T}_{02}^{F,\{3\}} \rangle_{\scriptscriptstyle \vec\xi,\theta_0^A}
    \left[ \dfrac{\Delta}{\Delta \xi_1} f (L_0,\bsxi,\theta_0^B)\right]
    -
    \langle {T}_{02}^{F,\{3\}} \rangle_{\scriptscriptstyle \vec\xi,\theta_0^B}
    \left[ \dfrac{\Delta}{\Delta \xi_1} f (L_0,\bsxi,\theta_0^A)\right]
  }{
    \langle {T}_{02}^{G,\{3\}} \rangle_{\scriptscriptstyle \vec\xi,\theta_0^A}\;
    \langle {T}_{02}^{F,\{3\}} \rangle_{\scriptscriptstyle \vec\xi,\theta_0^B}
    -
    \langle {T}_{02}^{G,\{3\}} \rangle_{\scriptscriptstyle \vec\xi,\theta_0^B}\;
    \langle {T}_{02}^{F,\{3\}} \rangle_{\scriptscriptstyle \vec\xi,\theta_0^A}
  }
\end{equation}
and
\begin{equation}\label{eq:ZF3}
  {\cal Z}^{\{3\}}_{F} = \dfrac{
    \langle {T}_{02}^{G,\{3\}} \rangle_{\scriptscriptstyle \vec\xi,\theta_0^B}
    \left[ \dfrac{\Delta}{\Delta \xi_1} f (L_0,\bsxi,\theta_0^A)\right]
    -
    \langle {T}_{02}^{G,\{3\}} \rangle_{\scriptscriptstyle \vec\xi,\theta_0^A}
    \left[ \dfrac{\Delta}{\Delta \xi_1} f (L_0,\bsxi,\theta_0^B)\right]
  }{
    \langle {T}_{02}^{G,\{3\}} \rangle_{\scriptscriptstyle \vec\xi,\theta_0^A}\;
    \langle {T}_{02}^{F,\{3\}} \rangle_{\scriptscriptstyle \vec\xi,\theta_0^B}
    -
    \langle {T}_{02}^{G,\{3\}} \rangle_{\scriptscriptstyle \vec\xi,\theta_0^B}\;
    \langle {T}_{02}^{F,\{3\}} \rangle_{\scriptscriptstyle \vec\xi,\theta_0^A}
  }
\end{equation}
for the triplet case.
The numerical results at fixed bare parameters
are collected in Table~\ref{tab:zetas} of 
Appendix~\ref{App:tables}. At $L_0/a=6$,  
${\cal Z}_G^{\{6\}}$, ${\cal Z}_F^{\{6\}}$, ${\cal Z}_G^{\{3\}}$ and
${\cal Z}_F^{\{3\}}$ have relative
errors of about $3\%$, $2\%$, $4\%$ and $3\%$ respectively.
At $L_0/a=4$ and $L_0/a=8$, the relative errors are about a half and twice
those figures, respectively. The precision changes by about $30\%-40\%$ in
the interval of $\beta$ explored in this study, the smaller the $\beta$ the
noisier the results. The errors quoted are statistical only, the others
being negligible with respect to them.

\subsection{Perturbative improvement}
\label{ssec:Perturbative improvement}
We have then considered the following perturbatively improved definition of
the renormalization constants,
\begin{equation}
    {\cal Z}_X^{\{i\}}(g_0^2, a/L_0) \rightarrow {\cal Z}_X^{\{i\}}(g_0^2, a/L_0)
    \dfrac{Z_X^{\{i\}(0)} + g_0^2\, Z_X^{\{i\}(1)}}
    {{\cal Z}_X^{\{i\}(0)}(a/L_0) + g_0^2\, {\cal Z}_X^{\{i\}(1)}(a/L_0)}\,,
    \label{eq:ZTptimpr}
\end{equation}
with $X=F,G$ for the fermionic and gluonic cases, so that discretization
effects up to one-loop order are subtracted to all orders in the lattice 
spacing. The determination of the tree-level and one-loop order 
perturbative coefficients at fixed $L_0/a$,
${\cal Z}_X^{\{i\}(0)}$ and ${\cal Z}_X^{\{i\}(1)}$, and in the
$1/L_0\to0$ limit, $Z_X^{\{i\}(0)}$ and $Z_X^{\{i\}(1)}$, is described in 
Appendix~\ref{app:appPT} where the tables with the numerical results are
also reported.

\subsection{The $1/L_0\rightarrow 0$ limit}\label{ssec:zeroT_extrap} 
Building on Eq.~\eqref{eq:aTto0_theory}, we have extrapolated each renormalization
constant to the $1/L_0\rightarrow 0$
limit by means of a global fit of the results at all values of the bare
parameters to the following generic fit function
\begin{equation}
  {\cal  Z}_X^{\{i\}}(g_0^2,a/L_0) = 1 +Z_X^{\{i\}(1)}g_0^2 
    + c_{X,2}^{\{i\}}\,g_0^4 +c_{X,3}^{\{i\}}\,g_0^6
    +\!\!\!\! \sum_{j\geq1,k\geq2} d_{X,jk}^{\{i\}} \Big(\dfrac{a}{L_0}\Big)^j g_0^{2k}\,,
    \label{eq:global_fit}
\end{equation}
where extrapolated results and $L_0$-dependent discretization effects
are parametrized by polynomials in the bare coupling squared. At
vanishing $1/L_0$, we have enforced the known perturbative results to one-loop
order~\cite{Caracciolo:1988hc,Caracciolo:1989bu,Caracciolo:1989pt,Caracciolo:1991cp,Caracciolo:1991vc,DallaBrida:2020gux}, that can be found in Appendix~\ref{app:appPT}.
We then allow for up to two terms, at $O(g_0^4)$ and
$O(g_0^6)$, whose respective coefficients $c_{X,2}^{\{i\}}$ and 
$c_{X,3}^{\{i\}}$ result from fitting non-perturbative data together
with the other coefficients $d_{X,jk}^{\{i\}}$ parameterizing discretization
effects. For the latter, the one-loop improvement of Eq.~\eqref{eq:ZTptimpr}
guarantees that terms with $k=0,1$ are not present.

In order to assess the stability of the $1/L_0\rightarrow 0$
extrapolation, we have considered several extrapolation procedures differing
by the terms included in the fit function in Eq.~\eqref{eq:global_fit},
and by either including or excluding the $L_0/a=4$ points in the fitted
dataset.
The results are always perfectly compatible among each other within error
bars. This study is reported in
Appendix~\ref{app:appExtrap}, while in the following we summarize the
final results for each renormalization constant. The complete covariance
matrix among the fit parameters entering the non-perturbative parametrizations
can be found in Table~\ref{tab:ZT_cov} of Appendix~\ref{app:appExtrap}.

\subsection{The sextet  case: $Z^{\{6\}}_{G}$ and $Z^{\{6\}}_{F}$}
\label{ssec:The_sextet_case}
\begin{figure*}[t]
\begin{center}
\begin{minipage}{0.5\columnwidth}
\includegraphics[width=0.95\textwidth]{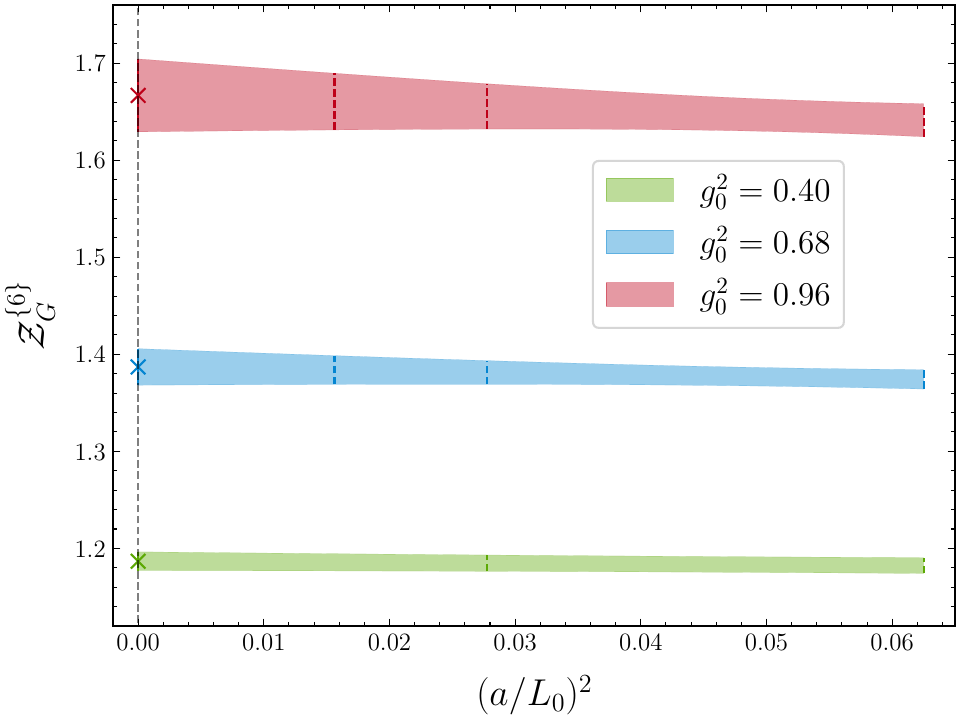}
\end{minipage}%
\begin{minipage}{0.5\columnwidth}
\includegraphics[width=0.95\textwidth]{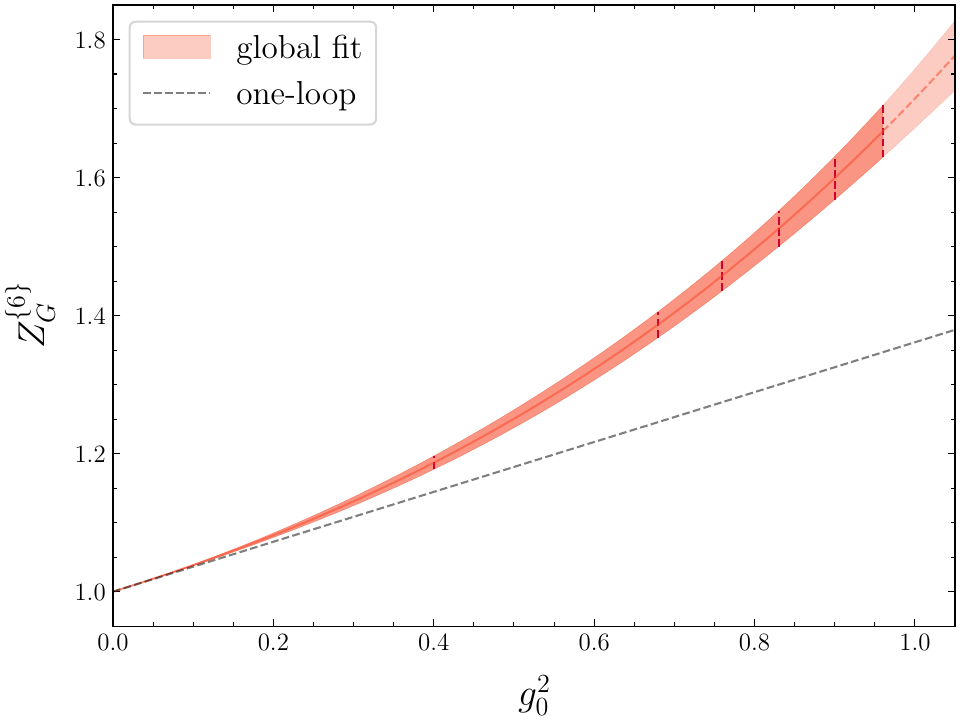}
\end{minipage}
    \caption{Left: representation of the approach to the $1/L_0\to0$ limit
    of the fit function ${\cal Z}^{\{6\}}_{G}(g_0^2, a/L_0)$ at three
    representative values of $g_0^2$. The values at $L_0/a=4,6,8$ are
    featured by dashed lines, while the limit is marked by a cross.
    Right: the red curve and its error band represent the parametrization
    of non-perturbative results for $Z^{\{6\}}_{G}(g_0^2)$. 
    The black line is the perturbative result to one-loop order.}
    \label{fig:ZG6_final}
\end{center}
\end{figure*}

\begin{figure*}[t]
\begin{center}
\begin{minipage}{0.5\columnwidth}
\includegraphics[width=0.95\textwidth]{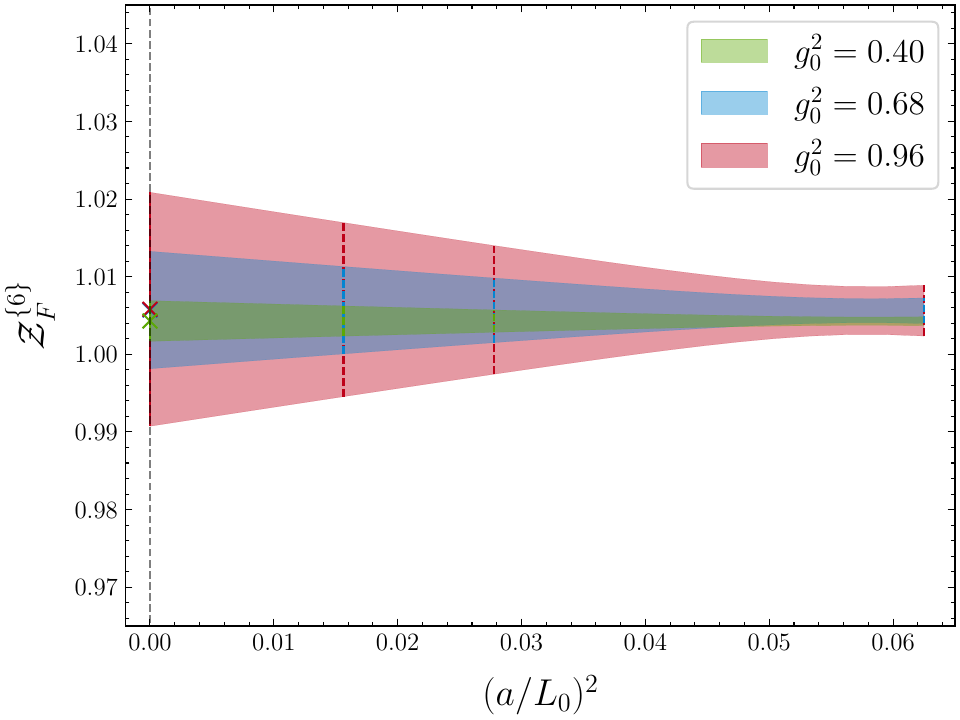}
\end{minipage}%
\begin{minipage}{0.5\columnwidth}
\includegraphics[width=0.95\textwidth]{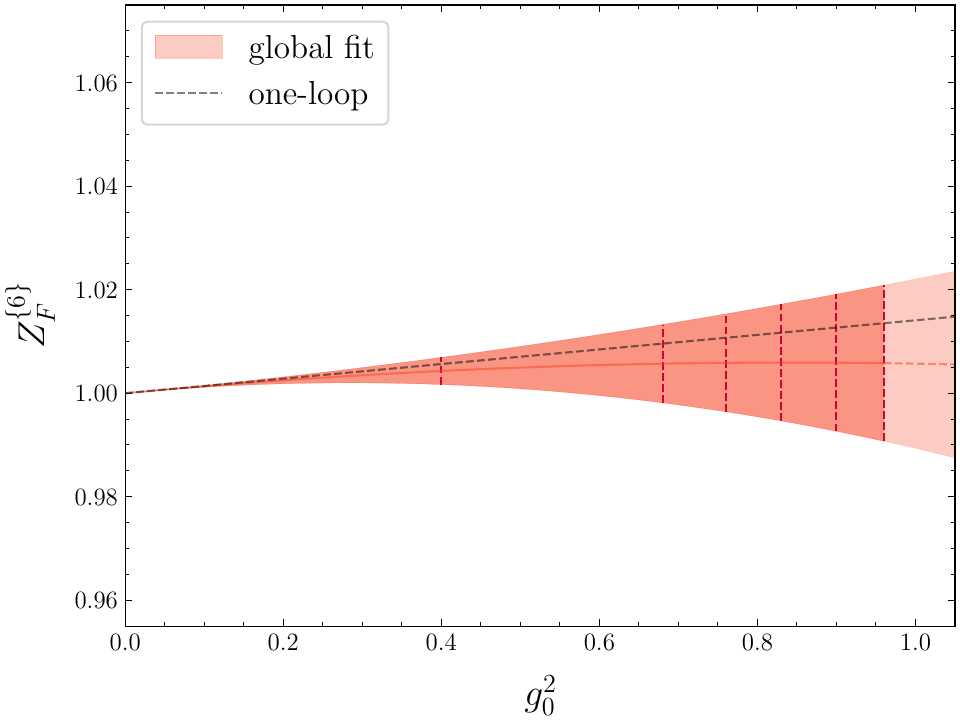}
\end{minipage}
    \caption{Left: representation of the approach to the $1/L_0\to0$ limit
    of the fit function ${\cal Z}^{\{6\}}_{F}(g_0^2, a/L_0)$ at three
    representative values of $g_0^2$. The values at $L_0/a=4,6,8$ are
    featured by dashed lines, while the limit is marked by a cross.
    Right: the red curve and its error band represent the parametrization
    of non-perturbative results for $Z^{\{6\}}_{F}(g_0^2)$. 
    The black line is the perturbative result to one-loop order.}
    \label{fig:ZF6_final}
\end{center}
\end{figure*}
The best result for the one-loop improved sextet gluonic 
renormalization constant $Z^{\{6\}}_{G}$ is obtained
by fitting all data, i.e. at $L_0/a=4,6,8$ and for all
$\beta$-values simulated, to the function in
Eq.~\eqref{eq:global_fit} with 
$c_{G,2}^{\{6\}}, c_{G,3}^{\{6\}}, d_{G,22}^{\{6\}}$ 
as non-vanishing coefficients. This fit has 
$\chi^2/\chi^2_{\rm exp}=0.94$ (with $\chi^2_{\rm exp}$ defined as in 
Ref.~\cite{Bruno:2022mfy}), and it has a very smooth approach to the 
$1/L_0\to0$ limit, as shown in the left panel of Figure~\ref{fig:ZG6_final}. 
In that limit, the fit leads to the parametrization
\begin{equation}
    Z^{\{6\}}_{G}(g_0^2) = 1 +Z^{\{6\}(1)}_{G}\,g_0^2
    +c_{G,2}^{\{6\}}\,g_0^4 +c_{G,3}^{\{6\}}\,g_0^6\,,
    \label{eq:ZG6_np_final}
\end{equation}
where $Z^{\{6\}(1)}_{G} = 0.361285(8)$ is the one-loop perturbative
value, while the fit fixes the coefficients $c_{G,2}^{\{6\}} = 0.20(10)$,
$c_{G,3}^{\{6\}} = 0.15(11)$ with off-diagonal entry of the covariance 
matrix, normalized to the product of the errors, equal to $-0.93$. This 
parametrization holds in the interval $0\leq g_0^2\leq 0.96$ of bare 
coupling constant squared, and the corresponding curve with its error band 
is represented in the right panel of Figure~\ref{fig:ZG6_final}. 
The results
at some selected values of $g_0^2$, chosen to be close to those actually
simulated in the interval $0.4\leq g_0^2\leq 0.96$, are represented as well 
in the figure by dashed lines, with their values being collected in Table~\ref{tab:Znp}. 
The relative error attached to the parametrization in 
Eq.~\eqref{eq:ZG6_np_final} increases quadratically for $g_0^2\leq0.2$ 
up to $0.3\%$, and then roughly linearly for $0.2\leq g_0^2\leq0.96$ up to
$2.3\%$. A clear difference with the one-loop perturbative result is
observed for  $g_0^2\geq 0.4$ or so. This is consistent with the results
obtained in the SU($3$) Yang--Mills theory~\cite{Giusti:2015daa,Giusti:2016iqr}.

The best fit of the sextet fermionic renormalization constant  
$Z^{\{6\}}_{F}$ has been carried out by considering $c_{F,2}^{\{6\}}$ 
and $d_{F,22}^{\{6\}}$ as non-vanishing coefficients in the function
in Eq.~\eqref{eq:global_fit}. We have included non-perturbative results
at $L_0/a=4,6,8$ for all $\beta$-values considered. The fit results
in a $\chi^2/\chi^2_{\rm exp}=1.23$, the $1/L_0\to0$ limit is approached
very smoothly (see left panel of Figure~\ref{fig:ZF6_final}), and the
final parametrization in that limit reads
\begin{equation}
    Z^{\{6\}}_{F}(g_0^2) = 1 +Z^{\{6\}(1)}_{F}\,g_0^2 
    +c_{F,2}^{\{6\}}\,g_0^4\,,
    \label{eq:ZF6_np_final}
\end{equation}
where $Z^{\{6\}(1)}_{F}=0.014085(11)$ is the one-loop perturbative 
coefficient, while the fit fixes $c_{F,2}^{\{6\}} = -0.008(16)$.
This parametrization holds in the interval $0\leq g_0^2\leq 0.96$ of bare 
coupling squared, and the corresponding curve with error band is 
represented on the right panel of Figure~\ref{fig:ZF6_final}. The results
at some selected values of $g_0^2$, chosen to be close to those actually
simulated in the interval $0.4\leq g_0^2\leq 0.96$, are represented as well
by dashed lines and their values are collected in Table~\ref{tab:Znp}. In the range
$0\leq g_0^2\leq 0.96$, the relative error attached to the 
parametrization in Eq.~\eqref{eq:ZF6_np_final} increases quadratically
up to to $1.5\%$. The non-perturbative results are well in agreement with
the one-loop perturbative ones in the entire interval.

\subsection{The triplet case: $Z^{\{3\}}_{G}$ and $Z^{\{3\}}_{F}$}
\label{ssec:The_triplet_case}

\begin{figure*}[t]
\begin{center}
\begin{minipage}{0.5\columnwidth}
\includegraphics[width=0.95\textwidth]{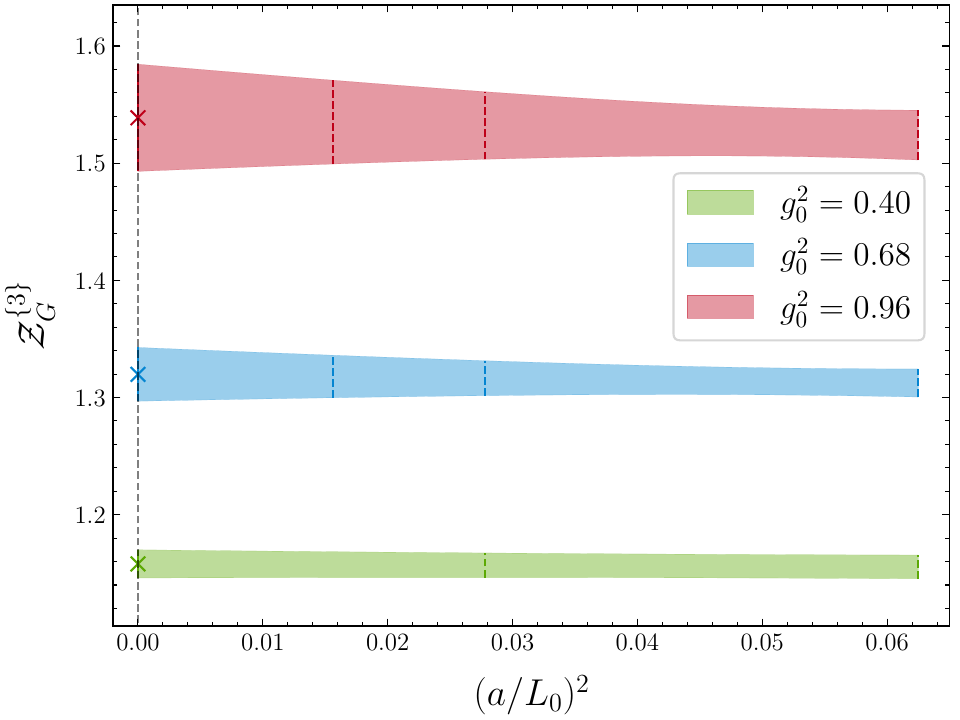}
\end{minipage}%
\begin{minipage}{0.5\columnwidth}
\includegraphics[width=0.95\textwidth]{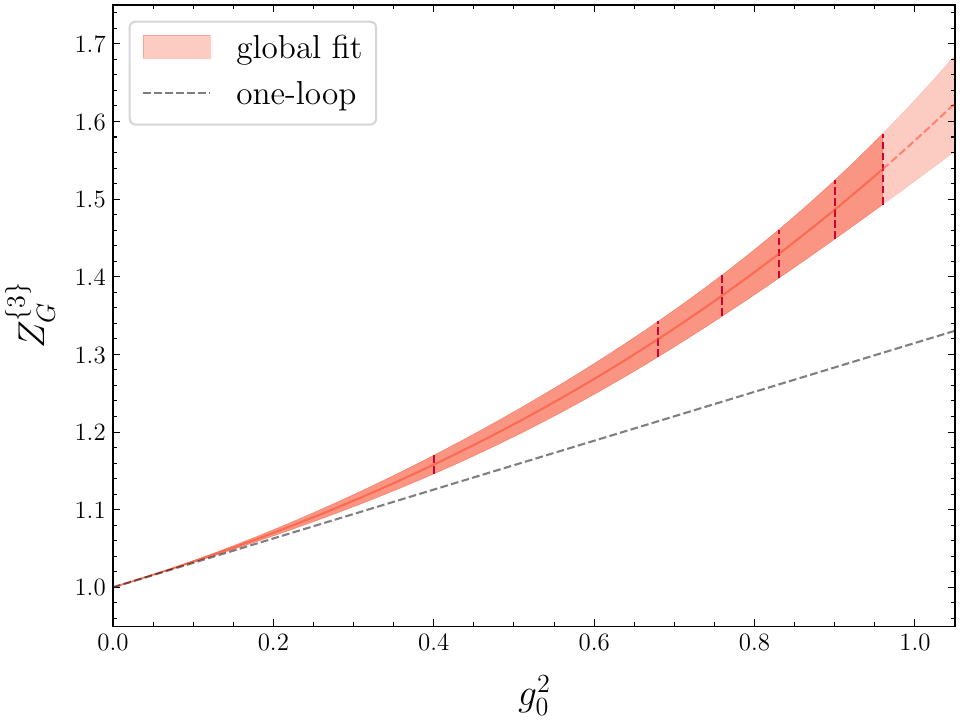}
\end{minipage}
    \caption{Left: representation of the approach to the $1/L_0\to0$ limit
    of the fit function ${\cal Z}^{\{3\}}_{G}(g_0^2, a/L_0)$ at three
    representative values of $g_0^2$. The values at $L_0/a=4,6,8$ are
    featured by dashed lines, while the limit is marked by a cross.
    Right: the red curve and its error band represent the parametrization
    of non-perturbative results for $Z^{\{3\}}_{G}(g_0^2)$. 
    The black line is the perturbative result to one-loop order.}
    \label{fig:ZG3_final}
\end{center}
\end{figure*}

\begin{figure*}[t]
\begin{center}
\begin{minipage}{0.5\columnwidth}
\includegraphics[width=0.95\textwidth]{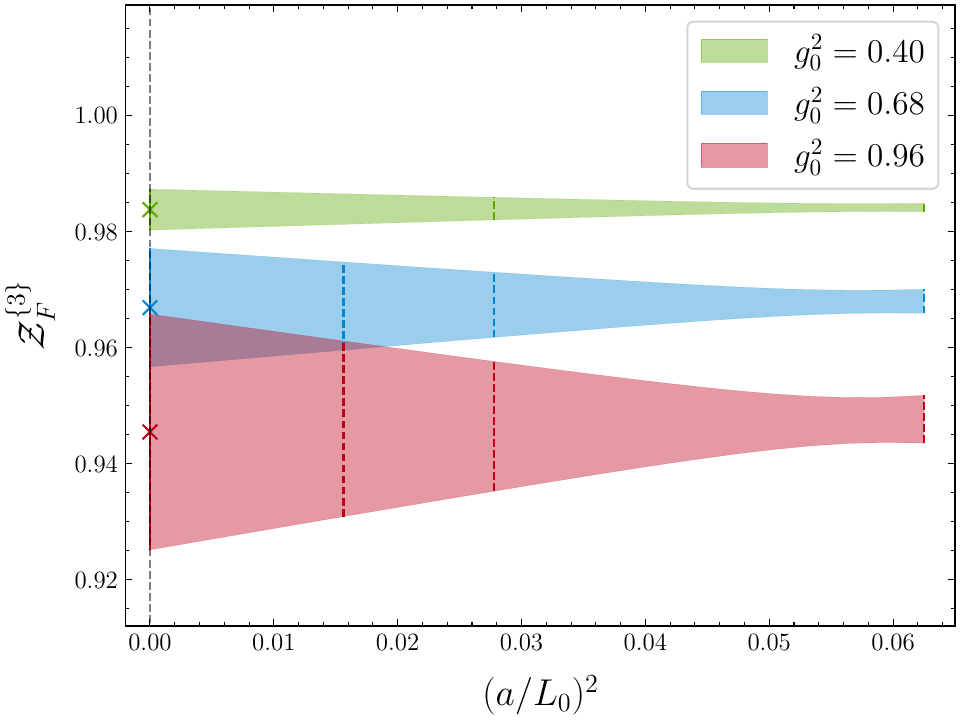}
\end{minipage}%
\begin{minipage}{0.5\columnwidth}
\includegraphics[width=0.95\textwidth]{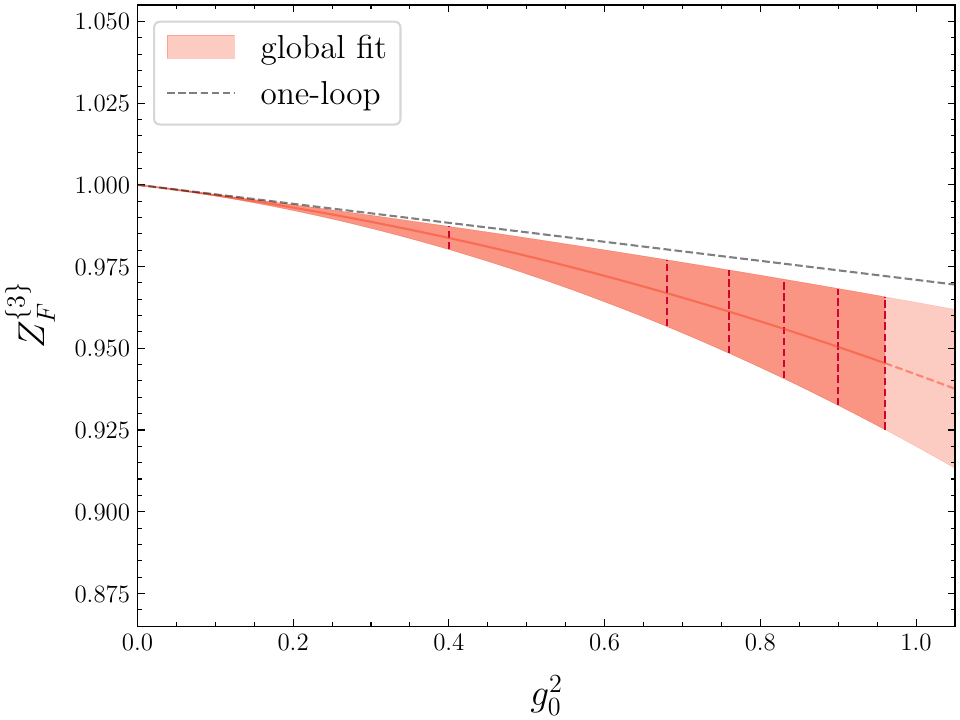}
\end{minipage}
    \caption{Left: representation of the approach to the $1/L_0\to0$ limit
    of the fit function ${\cal Z}^{\{3\}}_{F}(g_0^2, a/L_0)$ at three
    representative values of $g_0^2$. The values at $L_0/a=4,6,8$ are
    featured by dashed lines, while the limit is marked by a cross.
    Right: the red curve and its error band represent the parametrization
    of non-perturbative results for $Z^{\{3\}}_{F}(g_0^2)$. 
    The black line is the perturbative result to one-loop order.}
    \label{fig:ZF3_final}
\end{center}
\end{figure*}
The renormalization constants for the triplet representation of the 
energy-momentum tensor have been fitted by using the same procedures
described for the sextet. For the gluonic case, the fit has a 
$\chi^2/\chi^2_{\rm exp}=1.13$ and it leads to the following 
parametrization
\begin{equation}
    Z^{\{3\}}_{G}(g_0^2) = 1 +Z^{\{3\}(1)}_{G}\,g_0^2 
    +c_{G,2}^{\{3\}}\,g_0^4
    +c_{G,3}^{\{3\}}\,g_0^6\,,
    \label{eq:ZG3_np_final}
\end{equation}
where $Z^{\{3\}(1)}_{G}=0.31465(6)$ is the one-loop perturbative result,
while the fit fixes $c_{G,2}^{\{3\}} = 0.16(13)$ and 
$c_{G,3}^{\{3\}} = 0.10(14)$ with off-diagonal entry of the covariance 
matrix, normalized to the product of the errors, equal to $-0.93$.
The $1/L_0\to0$ extrapolation is shown in the left panel of
Figure~\ref{fig:ZG3_final}.
The fit for the triplet fermionic case leads to the parametrization
\begin{equation}
    Z^{\{3\}}_{F}(g_0^2) = 1 +Z^{\{3\}(1)}_{F}\,g_0^2 
    +c_{F,2}^{\{3\}}\,g_0^4\,,
    \label{eq:ZF3_np_final}
\end{equation}
where $Z^{\{3\}(1)}_{F}=-0.02907(27)$, while $c_{F,2}^{\{3\}} = -0.029(22)$
is fixed by the fit with $\chi^2/\chi^2_{\rm exp}=0.98$. The $1/L_0\to0$
extrapolation is shown in the left panel of Figure~\ref{fig:ZF3_final}.
Both parametrizations hold in the interval $0\leq g_0^2\leq 0.96$ of bare 
coupling squared, and the corresponding curves with error bands are
represented in the right panels of Figures~\ref{fig:ZG3_final} 
and~\ref{fig:ZF3_final}. The results at some selected values of $g_0^2$,
chosen to be close to those actually simulated in the interval 
$0.4\leq g_0^2\leq 0.96$, are represented as well in the figures by dashed lines, and their
values are reported in Table~\ref{tab:Znp}. The relative uncertainties are
about $30\%$ larger than the corresponding ones for the sextet 
representation, a fact that can be traced back to the
larger statistical fluctuations of the bare matrix elements of the 
energy-momentum tensor in the triplet representation. 
The comparison with the perturbative results to one-loop order leads to
similar qualitative conclusions as for the sextet case: while 
$Z^{\{3\}}_{G}$ differs from the one-loop perturbative result for
$g_0^2\geq 0.4$ or so, $Z^{\{3\}}_{F}$ shows a substantial agreement with
the one-loop perturbative result.

\begin{table}[t]
    \centering
    \begin{tabular}{|c|c|c|c|c|}
\hline
$g_0^2$ \rule[-6pt]{0pt}{20pt} & $Z_{G}^{\{6\}}$ & $Z_{F}^{\{6\}}$ & $Z_{G}^{\{3\}}$ & $Z_{F}^{\{3\}}$ \\
\hline
 $0.40$ &     $1.187(10)$ &    $1.0043(26)$ &     $1.158(12)$ &      $0.984(4)$ \\
 $0.68$ &     $1.387(19)$ &      $1.006(8)$ &     $1.320(23)$ &     $0.967(10)$ \\
 $0.76$ &     $1.458(22)$ &      $1.006(9)$ &     $1.376(27)$ &     $0.961(13)$ \\
 $0.83$ &     $1.526(26)$ &     $1.006(11)$ &       $1.43(3)$ &     $0.956(15)$ \\
 $0.90$ &       $1.60(3)$ &     $1.006(13)$ &       $1.49(4)$ &     $0.950(18)$ \\
 $0.96$ &       $1.67(4)$ &     $1.006(15)$ &       $1.54(5)$ &     $0.945(20)$ \\
\hline
\end{tabular}

    \caption{Non-perturbative results for the renormalization constants at
    some selected values of the bare coupling squared, obtained using the
    parametrizations presented in the main text.}
    \label{tab:Znp}
\end{table}

\section{Conclusions}
In this work we have established a fully non-perturbative definition of the
traceless components of the QCD energy-momentum tensor whose correlation
functions satisfy the appropriate Ward identities in the continuum limit.
The crucial step forward to reach this goal has been the non-perturbative
determination of the renormalization constants of the triplet and the
sextet components of the tensor.
The key theoretical ingredients have been the use of shifted boundary conditions
for quark and gluon fields in the presence of a compact temporal direction,
supplemented by a twist phase for the quark fields~\cite{Giusti:2015daa,DallaBrida:2020gux}.
In this setup,
exact Ward identities for one-point functions can be derived which, once enforced
on the lattice up to discretization effects which vanish power-like in the continuum limit,
define non-perturbative renormalization conditions for the energy-momentum tensor
that can be efficiently imposed numerically.

Our final results for the renormalization factors
$Z_{G}^{\{6\}}(g_0^2)$, $Z_{F}^{\{6\}}(g_0^2)$, $Z_{G}^{\{3\}}(g_0^2)$ and
$Z_{F}^{\{3\}}(g_0^2)$ are given in Eqs.~(\ref{eq:ZG6_np_final}),
(\ref{eq:ZF6_np_final}), (\ref{eq:ZG3_np_final}) and (\ref{eq:ZF3_np_final}),
respectively. They are parameterized by polynomials in the bare coupling constant
squared $g_0^2$,
holding in the interval $0\leq g_0^2\leq 0.96$. The percent-level
uncertainties are dominated by the statistical errors, thus indicating that
the precision can be further improved by increasing the Monte Carlo sampling.
The fermionic renormalization factors $Z_{F}^{\{6\}}(g_0^2)$ and $Z_{F}^{\{3\}}(g_0^2)$ are
very close to $1$, and they are compatible within the statistical accuracy
with the one-loop perturbation theory
results in the interval of $g_0^2$ explored. Conversely, the
gluonic renormalization constants $Z_{G}^{\{6\}}(g_0^2)$ and $Z_{G}^{\{3\}}(g_0^2)$
are significantly larger than the one-loop perturbative
results. Already at $g_0^2 = 0.4$, the smallest bare coupling constant
squared that we have simulated non-perturbatively, a $3\%$ deviation
(about 3 sigma in terms of the non-perturbative error)
from the one-loop results is observed. The deviation becomes $20\%$ or so
at the largest coupling simulated, $g_0^2=0.96$. A very similar behaviour
was observed in the SU($3$) pure gauge theory in
Refs.~\cite{Giusti:2015daa,Giusti:2016iqr}.

These findings demonstrate that the 
non-perturbative definition of the energy-momentum tensor is required at
values of $g_0^2$ relevant for lattice QCD simulations at zero
and finite temperature. In particular it is needed for phenomenological
applications where errors of a
few percent or less are required, e.g. the study of transport properties
of the QCD plasma. From a more theoretical
point of view, the results presented here pave the way for a 
non-perturbative matching of the energy-momentum tensor with the analogous
quantity defined at non-zero flow time~\cite{Luscher:2010iy,Suzuki:2013gza}.

\section*{Acknowledgments}
We acknowledge PRACE for awarding us access to the HPC system MareNostrum4 at the Barcelona Supercomputing
Center (Proposals n. 2018194651 and 2021240051) where some of the numerical results presented in this
letter have been obtained. We also thank CINECA for providing us with a very generous access
to Leonardo during the early phases of operations of the machine and for the computer time
allocated via the CINECA-INFN, CINECA-Bicocca agreements. The R\&D has been carried out on the PC
clusters Wilson and Knuth at Milano-Bicocca. We thank all these institutions for the technical
support. This work is (partially) supported by ICSC - Centro Nazionale di Ricerca in High
Performance Computing, Big Data and Quantum Computing, funded by European Union - NextGenerationEU.
The numerical simulations have been carried out implementing shifted boundary conditions on the
code \texttt{openQCD-1.6.0}~\cite{openQCD}.

\appendix

\section{QCD lattice action\label{App:action}}
In this Appendix we summarize the lattice formulation of QCD used in this study.
Definitions not given here can be found in Ref.~\cite{DallaBrida:2020gux},
where a more extensive discussion is reported.
The lattice QCD action decomposes as
\begin{equation}\label{eq:LatAction}
S = S^G + S^F 
\end{equation}
with $S^G$ being the Yang–Mills action and $S^F$ the $O(a)$-improved Wilson fermion action.
For the gauge field we adopt the Wilson plaquette action
\begin{equation}
\label{eq:SG}
S^G = {1 \over g_0^2} \sum_x \sum_{\mu,\nu} {\rm Re}\,\Tr\big\{ \mathbf 1 - U_{\mu\nu}(x) \big\},
\end{equation}
where $g_0$ is the bare gauge coupling and $U_{\mu\nu}(x)$ denotes the standard plaquette
\begin{equation}
U_{\mu\nu}(x)=U_\mu(x)U_\nu(x+ a \hat{\mu})U_\mu^\dag(x+ a \hat{\nu})U_\nu^\dag(x)\; ,
\end{equation}
with $\hat \mu$ and $\hat \nu$ being the unit vectors in the $\mu$ and $\nu$ directions.
The fermionic contribution to the action is
\begin{equation}\label{eq:fermionicAction}
S^F = a^4 \sum_x \psibar(x) (D+M_0)\psi(x)\; , 
\end{equation}
where $M_0$ is the bare quark mass matrix, and $D$ is the O($a$)-improved Wilson–Dirac operator,
\begin{equation}
\label{eq:Dirac}
D = D_{\rm w} + a D_{\rm sw}\, .
\end{equation}
The massless Wilson–Dirac operator is defined as
\begin{equation}
D_{\rm w}
= \frac12 \big\{\dirac\mu(\nabstar\mu+\nab\mu)-a\nabstar\mu \nab\mu\big\}\; ,
\end{equation}
where $\dirac\mu$ are the Dirac matrices in Euclidean space satisfying $\{ \gamma_\mu,\gamma_\nu \} = 2 \delta_{\mu\nu} \mathbf 1$.
The forward and backward gauge covariant lattice derivatives act on the fermion fields according to
\ba
a \nab\mu \psi(x) & = & U_\mu(x)\psi(x+ a \hat{\mu})-\psi(x)\; ,\nonumber \\[0.25cm]
a \nabstar\mu \psi(x) & = & \psi(x) - U^\dag_\mu(x- a \hat{\mu})\psi(x - a \hat{\mu})\; .
\label{eq:fwd-nablas}
\ea
We also introduce the backward derivatives acting on $\psibar$ 
\ba
  \nonumber
  a  \psibar(x)\nabbar\mu &= &\psibar(x+a \hat{\mu})U^\dag_\mu(x) - \psibar(x)\; ,\\[0.25cm]
  a \psibar(x)\nabbarstar\mu & = &\psibar(x)-\psibar(x-a \hat{\mu})U_\mu(x-a \hat{\mu})\; ,
\ea
and the combinations 
\begin{equation}\label{eq:DbfL}
  \nabdbar\mu = \nab\mu - \nabbar\mu\; ,
  \qquad
  \nabdbarstar\mu = \nabstar\mu - \nabbarstar\mu\; ,
\end{equation}
that appear in the definition of the fermionic part of the energy-momentum tensor.

The $O(a)$-improvement term is the Sheikholeslami–Wohlert operator \cite{Sheikholeslami:1985ij},
\begin{equation}
\label{eq:DiracSW}
	D_{\rm sw}\psi(x) = c_{\rm sw}(g_0) {1 \over 4}
	\sigma_{\mu\nu} \widehat F_{\mu\nu}(x)\psi(x)\; ,
\end{equation}
where $\sigma_{\mu\nu} = \tfrac{i}{2}[\gamma_\mu,\gamma_\nu]$ and $\widehat F_{\mu\nu}(x)$ is the clover discretization of the
field-strength tensor. For historical reasons, the clover discretization $\widehat F_{\mu\nu}$ used in Eq.~\eqref{eq:DiracSW} is not
traceless, unlike the discretization adopted for the definition of the energy–momentum tensor. The clover term is
constructed from the plaquette loops
\begin{equation}
\label{eq:CloverFmunu}
\widehat F_{\mu\nu}(x) = {i\over 8a^2}\big\{Q_{\mu\nu}(x)-Q_{\nu\mu}(x)\big\}\; ,
\end{equation}
with
\begin{equation}
\begin{split}
	Q_{\mu\nu}(x) &= U_\mu(x)U_\nu(x+a \hat{\mu})U^\dag_\mu(x+a \hat{\nu})U^\dag_\nu(x)\\
	&+ U_\nu(x)U_\mu^\dag(x-a \hat{\mu}+a \hat{\nu})U^\dag_\nu(x-a \hat{\mu})U_\mu(x-a \hat{\mu})\\
	&+ U_\mu^\dag(x-a \hat{\mu})U_\nu^\dag(x-a \hat{\mu}-a \hat{\nu})U_\mu(x-a \hat{\mu}-a \hat{\nu})U_\nu(x-a \hat{\nu})\\
	&+ U_\nu^\dag(x-a \hat{\nu})U_\mu(x-a \hat{\nu})U_\nu(x+a \hat{\mu}-a \hat{\nu})U_\mu^\dag(x)\; .
\end{split}
\end{equation}
The improvement coefficient $c_{\rm sw}(g_0)$ is tuned non-perturbatively in order to cancel $O(a)$ effects in on-shell correlation
functions~\cite{Sheikholeslami:1985ij,Luscher:1996sc}, and is given by the following expression~\cite{Yamada:2004ja},
\begin{equation}\label{eq:cSW-W}
  c_{\rm sw} (g_0^2)=\dfrac
  {1 - 0.194785\, g_0^2 - 0.110781\, g_0^4 - 0.0230239\, g_0^6 + 0.137401\, g_0^8}
  {1-0.460685\, g_0^2} \, .
\end{equation}

\section{Data tables\label{App:tables}}

\begin{table}[ht!]
\centering
\begin{tabular}{|c|c|c|c|c|}
\hline
$\beta=6/g_0^2$ \rule[-8pt]{0pt}{22pt}
& $\corr{T_{01}^{G,\{6\}}}_{\scriptscriptstyle\vec\xi,\theta_0^A}\times 10^4$
& $\corr{T_{01}^{F,\{6\}}}_{\scriptscriptstyle\vec\xi,\theta_0^A}\times 10^4$
& $\corr{T_{02}^{G,\{3\}}}_{\scriptscriptstyle\vec\xi,\theta_0^A}\times 10^4$
& $\corr{T_{02}^{F,\{3\}}}_{\scriptscriptstyle\vec\xi,\theta_0^A}\times 10^4$ \\
\hline
\multicolumn{5}{|c|}{$L_0/a=4$} \\
\hline
 $6.3719$ &  $-23.81(6)$ & $-63.028(20)$ &  $-27.87(8)$ &  $-70.26(4)$ \\
 $6.7079$ &  $-24.65(5)$ & $-64.063(21)$ &  $-28.73(6)$ &  $-71.24(3)$ \\
 $7.0250$ &  $-25.29(5)$ & $-64.957(24)$ &  $-29.26(5)$ &  $-72.13(3)$ \\
 $7.3534$ &  $-25.95(6)$ & $-65.688(20)$ &  $-29.91(7)$ &  $-72.76(4)$ \\
 $7.6713$ &  $-26.41(5)$ & $-66.339(20)$ &  $-30.13(6)$ & $-73.332(27)$ \\
 $7.9794$ &  $-26.82(6)$ & $-66.870(18)$ &  $-30.65(8)$ & $-73.850(25)$ \\
 $8.3033$ &  $-27.29(4)$ & $-67.405(18)$ &  $-31.02(7)$ & $-74.312(23)$ \\
$15.0000$ &  $-30.78(9)$ & $-72.248(27)$ & $-33.75(12)$ &  $-79.01(4)$ \\
\hline
\multicolumn{5}{|c|}{$L_0/a=6$} \\
\hline
 $6.2735$ &   $-4.55(4)$ & $-11.317(20)$ &   $-5.21(5)$ &  $-12.35(3)$ \\
 $6.6050$ &   $-4.73(4)$ & $-11.476(16)$ &   $-5.41(4)$ & $-12.481(26)$ \\
 $6.9433$ &   $-4.95(3)$ & $-11.624(19)$ &   $-5.56(6)$ & $-12.667(29)$ \\
 $7.2618$ &   $-5.01(4)$ & $-11.778(17)$ &   $-5.50(6)$ & $-12.767(25)$ \\
 $7.5909$ &   $-5.14(4)$ & $-11.905(16)$ &   $-5.82(4)$ & $-12.884(21)$ \\
 $7.9091$ &   $-5.20(5)$ & $-11.989(13)$ &   $-5.82(5)$ & $-12.963(19)$ \\
 $8.2170$ & $-5.238(28)$ & $-12.090(15)$ &   $-5.82(5)$ & $-13.066(20)$ \\
 $8.5403$ &   $-5.41(4)$ & $-12.169(16)$ &   $-5.95(6)$ & $-13.178(19)$ \\
$15.0000$ &   $-6.02(4)$ & $-12.934(8)$ &   $-6.37(5)$ & $-13.902(11)$ \\
\hline
\multicolumn{5}{|c|}{$L_0/a=8$} \\
\hline
 $6.4680$ & $-1.478(20)$ & $-3.479(11)$ & $-1.730(29)$ & $-3.721(16)$ \\
 $6.7915$ & $-1.532(25)$ & $-3.512(10)$ &   $-1.74(3)$ & $-3.771(14)$ \\
 $7.1254$ & $-1.640(23)$ & $-3.573(11)$ &   $-1.73(3)$ & $-3.809(17)$ \\
 $7.4424$ & $-1.606(24)$ & $-3.640(10)$ & $-1.837(29)$ & $-3.815(14)$ \\
 $7.7723$ & $-1.629(22)$ &  $-3.657(9)$ &   $-1.83(3)$ & $-3.882(13)$ \\
 $8.0929$ & $-1.697(23)$ &  $-3.679(9)$ & $-1.914(26)$ & $-3.886(14)$ \\
 $8.4044$ & $-1.735(26)$ &  $-3.717(9)$ &   $-1.89(3)$ & $-3.908(13)$ \\
 $8.7325$ & $-1.699(23)$ &  $-3.743(9)$ &   $-1.92(4)$ & $-3.930(13)$ \\
\hline
\end{tabular}

\caption{Non-perturbative results of the sextet and triplet components
  of the bare energy-momentum tensor, determined on lattices with shifted
 boundary conditions specified by
$\vec\xi=(1,0,0)$ and $\theta_0=\theta_0^A = 0$.}  
  \label{tab:tensor_thA}
\end{table}

\begin{table}[ht!]
\centering
\begin{tabular}{|c|c|c|c|c|}
\hline
$\beta=6/g_0^2$ \rule[-8pt]{0pt}{22pt}
& $\corr{T_{01}^{G,\{6\}}}_{\scriptscriptstyle\vec\xi,\theta_0^B}\times 10^4$
& $\corr{T_{01}^{F,\{6\}}}_{\scriptscriptstyle\vec\xi,\theta_0^B}\times 10^4$
& $\corr{T_{02}^{G,\{3\}}}_{\scriptscriptstyle\vec\xi,\theta_0^B}\times 10^4$
& $\corr{T_{02}^{F,\{3\}}}_{\scriptscriptstyle\vec\xi,\theta_0^B}\times 10^4$ \\
\hline
\multicolumn{5}{|c|}{$L_0/a=4$} \\
\hline
 $6.3719$ &  $-21.76(8)$ &  $-40.05(4)$ & $-24.92(11)$ &  $-44.71(5)$ \\
 $6.7079$ &  $-22.65(7)$ &  $-40.65(3)$ & $-25.82(11)$ &  $-45.30(4)$ \\
 $7.0250$ & $-23.45(10)$ &  $-41.21(3)$ & $-26.48(10)$ &  $-45.79(5)$ \\
 $7.3534$ & $-24.03(11)$ &  $-41.52(3)$ & $-27.28(12)$ &  $-46.23(4)$ \\
 $7.6713$ &  $-24.55(9)$ &  $-41.92(3)$ & $-27.69(13)$ &  $-46.47(4)$ \\
 $7.9794$ & $-24.99(10)$ &  $-42.21(3)$ & $-28.08(13)$ &  $-46.71(4)$ \\
 $8.3033$ & $-25.57(13)$ &  $-42.53(3)$ & $-28.55(15)$ &  $-46.99(4)$ \\
$15.0000$ & $-29.73(10)$ &  $-45.38(3)$ & $-32.02(15)$ &  $-49.57(4)$ \\
\hline
\multicolumn{5}{|c|}{$L_0/a=6$} \\
\hline
 $6.2735$ & $-4.116(24)$ &  $-7.276(9)$ & $-4.687(29)$ & $-7.989(13)$ \\
 $6.6050$ & $-4.341(22)$ &  $-7.355(8)$ &   $-4.76(3)$ & $-8.076(12)$ \\
 $6.9433$ & $-4.438(24)$ &  $-7.465(8)$ &   $-5.04(3)$ & $-8.157(13)$ \\
 $7.2618$ & $-4.622(29)$ & $-7.541(11)$ &   $-5.09(4)$ & $-8.223(15)$ \\
 $7.5909$ &   $-4.72(3)$ & $-7.615(11)$ &   $-5.22(5)$ & $-8.279(15)$ \\
 $7.9091$ &   $-4.84(3)$ & $-7.655(10)$ &   $-5.29(5)$ & $-8.313(15)$ \\
 $8.2170$ &   $-4.98(4)$ & $-7.711(10)$ &   $-5.35(4)$ & $-8.345(14)$ \\
 $8.5403$ &   $-5.00(3)$ & $-7.748(10)$ &   $-5.52(5)$ & $-8.407(14)$ \\
$15.0000$ &   $-5.77(4)$ &  $-8.177(9)$ &   $-6.14(5)$ & $-8.757(12)$ \\
\hline
\multicolumn{5}{|c|}{$L_0/a=8$} \\
\hline
 $6.4680$ & $-1.329(14)$ &  $-2.251(6)$ & $-1.468(20)$ &  $-2.405(8)$ \\
 $6.7915$ & $-1.436(16)$ &  $-2.275(6)$ & $-1.531(20)$ &  $-2.432(8)$ \\
 $7.1254$ & $-1.469(15)$ &  $-2.307(6)$ & $-1.587(22)$ &  $-2.460(8)$ \\
 $7.4424$ & $-1.487(16)$ &  $-2.331(5)$ & $-1.582(21)$ &  $-2.479(8)$ \\
 $7.7723$ & $-1.532(16)$ &  $-2.355(5)$ & $-1.656(21)$ &  $-2.492(8)$ \\
 $8.0929$ & $-1.540(16)$ &  $-2.368(5)$ & $-1.673(22)$ &  $-2.508(8)$ \\
 $8.4044$ & $-1.579(18)$ &  $-2.375(5)$ & $-1.707(23)$ &  $-2.536(7)$ \\
 $8.7325$ & $-1.626(18)$ &  $-2.386(5)$ & $-1.728(24)$ &  $-2.536(7)$ \\
\hline
\end{tabular}

\caption{Non-perturbative results of the sextet and triplet components
  of the bare energy-momentum tensor, determined on lattices with
  shifted boundary conditions specified by
$\vec\xi=(1,0,0)$ and $\theta_0=\theta_0^B = 3\pi/10$.}   
  \label{tab:tensor_thB}
\end{table}

\begin{table}[ht!]
\centering
\begin{tabular}{|c|c|c|c|c|c|}
\hline
\multirow[c]{2}{*}{$\beta=6/g_0^2$}
& \multirow[c]{2}{*}{$\dfrac{\Delta}{\Delta\xi_1} f \times 10^4$}
&  \multicolumn{3}{c|}{
$\dfrac{\Delta}{\Delta\xi_1}
\corr{V_0}_{\scriptscriptstyle\vec\xi,\theta_0} \times 10^3$ 
\rule[-12pt]{0pt}{30pt}}
&  \multirow[c]{2}{*}{${\cal V}^{AB}_{0,1}\times 10^4$} \\
\cline{3-5}
& & $\theta_0=0.3164$ & $\theta_0=0.6664$ &
$\theta_0=0.8878$ &  \rule[-2pt]{0pt}{14pt} \\
\hline
\multicolumn{6}{|c|}{$L_0/a=4$} \\
\hline
 $6.3719$ & $142.32(9)$ & $11.891(16)$ & $23.951(16)$ & $30.554(15)$ & $39.941(20)$ \\
 $6.7079$ & $143.80(7)$ & $12.011(16)$ & $24.224(14)$ & $30.854(17)$ & $40.365(20)$ \\
 $7.0250$ & $144.93(6)$ & $12.102(16)$ & $24.449(18)$ & $31.116(17)$ & $40.711(22)$ \\
 $7.3534$ & $146.07(6)$ & $12.210(15)$ & $24.636(17)$ & $31.386(15)$ & $41.049(20)$ \\
 $7.6713$ & $146.87(6)$ & $12.301(14)$ & $24.795(16)$ & $31.554(16)$ & $41.313(19)$ \\
 $7.9794$ & $147.55(6)$ & $12.346(13)$ & $24.910(15)$ & $31.759(14)$ & $41.514(18)$ \\
 $8.3033$ & $148.29(5)$ & $12.412(15)$ & $25.060(13)$ & $31.902(13)$ & $41.739(18)$ \\
$15.0000$ & $154.64(6)$ & $13.060(13)$ & $26.334(13)$ & $33.537(12)$ & $43.882(16)$ \\
\hline
\multicolumn{6}{|c|}{$L_0/a=6$} \\
\hline
 $6.2735$ & $21.14(7)$ & $2.412(10)$ & $4.932(14)$ & $6.306(14)$ & $5.463(10)$ \\
 $6.6050$ & $21.38(6)$ & $2.420(13)$ & $4.951(13)$ & $6.349(12)$ & $5.487(11)$ \\
 $6.9433$ & $21.50(5)$ & $2.454(13)$ & $4.967(13)$ & $6.384(14)$ & $5.526(11)$ \\
 $7.2618$ & $21.58(5)$ & $2.447(12)$ & $4.998(12)$ & $6.420(14)$ & $5.545(11)$ \\
 $7.5909$ & $21.72(5)$ & $2.470(13)$ & $4.993(12)$ & $6.423(12)$ & $5.558(11)$ \\
 $7.9091$ & $21.80(5)$ & $2.491(12)$ & $5.031(13)$ & $6.434(13)$ & $5.593(11)$ \\
 $8.2170$ & $21.96(4)$ & $2.497(13)$ & $5.065(12)$ & $6.468(12)$ & $5.621(10)$ \\
 $8.5403$ & $21.92(4)$ & $2.480(9)$ & $5.068(12)$ & $6.476(13)$ & $5.613(9)$ \\
$15.0000$ & $22.45(8)$ & $2.555(13)$ & $5.218(12)$ & $6.681(12)$ & $5.784(11)$ \\
\hline
\multicolumn{6}{|c|}{$L_0/a=8$} \\
\hline
 $6.4680$ &  $6.30(4)$ & $0.926(15)$ & $1.890(15)$ & $2.429(16)$ & $1.573(10)$ \\
 $6.7915$ &  $6.32(4)$ & $0.917(16)$ & $1.898(15)$ & $2.458(15)$ & $1.576(10)$ \\
 $7.1254$ & $6.326(23)$ & $0.936(15)$ & $1.922(16)$ & $2.414(15)$ & $1.587(10)$ \\
 $7.4424$ & $6.403(29)$ & $0.922(15)$ & $1.924(14)$ & $2.449(16)$ & $1.587(10)$ \\
 $7.7723$ & $6.389(26)$ & $0.959(16)$ & $1.922(14)$ & $2.467(17)$ & $1.608(10)$ \\
 $8.0929$ & $6.388(25)$ & $0.956(14)$ & $1.918(15)$ & $2.471(15)$ & $1.606(9)$ \\
 $8.4044$ & $6.443(23)$ & $0.936(14)$ & $1.933(15)$ & $2.477(15)$ & $1.602(9)$ \\
 $8.7325$ & $6.448(26)$ & $0.936(15)$ & $1.950(14)$ & $2.453(16)$ & $1.604(9)$ \\
\hline
\end{tabular}

\caption{Second column: non-perturbative results for the discrete 
derivative in the shift of the free-energy density. 
The boundary conditions are specified by the shift 
$\vec\xi=(1,0,0)$ and the fermionic twist phase $\theta_0=\theta_0^A=0$. Third to fifth columns: non-perturbative results for the discrete
derivative of the conserved vector current, computed at the values of 
$\theta_0$ required for the determination of the quantity 
${\cal V}_{0,1}^{AB}$ as defined in Eq.~\eqref{eq:V_{0,k}^{AB}}. The
results for the latter are reported in the last column.}
\label{tab:dfdv}
\end{table}

\begin{table}[ht!]
\centering
\begin{tabular}{|c|c|c|c|c|}
\hline
$\beta=6/g_0^2$ \rule[-6pt]{0pt}{20pt}
& ${\cal Z}_G^{\{6\}}$ & ${\cal Z}_F^{\{6\}}$
& ${\cal Z}_G^{\{3\}}$ & ${\cal Z}_F^{\{3\}}$ \\
\hline
\multicolumn{5}{|c|}{$L_0/a=4$} \\
\hline
 $6.3719$ &  $1.799(28)$ &  $1.578(11)$ &    $1.64(3)$ &  $1.374(14)$ \\
 $6.7079$ &  $1.738(22)$ &   $1.576(9)$ &  $1.588(29)$ &  $1.378(12)$ \\
 $7.0250$ &  $1.658(25)$ &  $1.585(10)$ &  $1.545(26)$ &  $1.383(11)$ \\
 $7.3534$ &  $1.664(26)$ &  $1.566(11)$ &  $1.476(25)$ &  $1.401(11)$ \\
 $7.6713$ &  $1.622(22)$ &   $1.568(9)$ &  $1.453(27)$ &  $1.406(11)$ \\
 $7.9794$ &  $1.600(24)$ &  $1.565(10)$ &  $1.461(25)$ &  $1.392(11)$ \\
 $8.3033$ &  $1.556(29)$ &  $1.570(12)$ &  $1.429(26)$ &  $1.399(11)$ \\
$15.0000$ &  $1.311(15)$ &   $1.582(7)$ &  $1.266(23)$ &  $1.416(11)$ \\
\hline
\multicolumn{5}{|c|}{$L_0/a=6$} \\
\hline
 $6.2735$ &    $1.76(6)$ &  $1.162(28)$ &    $1.52(6)$ &    $1.07(3)$ \\
 $6.6050$ &    $1.68(5)$ &  $1.172(23)$ &    $1.64(7)$ &    $1.00(3)$ \\
 $6.9433$ &    $1.72(5)$ &  $1.116(24)$ &    $1.46(5)$ &  $1.057(29)$ \\
 $7.2618$ &    $1.57(5)$ &  $1.165(24)$ &    $1.38(6)$ &  $1.096(29)$ \\
 $7.5909$ &    $1.58(5)$ &  $1.142(24)$ &    $1.49(6)$ &  $1.011(29)$ \\
 $7.9091$ &    $1.50(5)$ &  $1.168(26)$ &    $1.43(6)$ &  $1.039(28)$ \\
 $8.2170$ &    $1.43(4)$ &  $1.199(19)$ &    $1.41(5)$ &  $1.051(25)$ \\
 $8.5403$ &    $1.51(4)$ &  $1.129(23)$ &    $1.35(5)$ &  $1.053(25)$ \\
$15.0000$ &    $1.26(4)$ &  $1.150(17)$ &    $1.19(4)$ &  $1.070(18)$ \\
\hline
\multicolumn{5}{|c|}{$L_0/a=8$} \\
\hline
 $6.4680$ &   $1.75(11)$ &    $1.07(5)$ &   $1.87(17)$ &    $0.82(9)$ \\
 $6.7915$ &    $1.46(9)$ &    $1.16(5)$ &   $1.64(12)$ &    $0.92(7)$ \\
 $7.1254$ &    $1.59(9)$ &    $1.04(5)$ &   $1.40(10)$ &    $1.03(6)$ \\
 $7.4424$ &    $1.56(9)$ &    $1.07(5)$ &   $1.69(13)$ &    $0.87(7)$ \\
 $7.7723$ &    $1.38(7)$ &    $1.13(4)$ &   $1.41(10)$ &    $0.98(6)$ \\
 $8.0929$ &    $1.50(8)$ &    $1.04(4)$ &   $1.51(10)$ &    $0.90(6)$ \\
 $8.4044$ &    $1.54(9)$ &    $1.02(5)$ &   $1.37(10)$ &    $0.99(5)$ \\
 $8.7325$ &    $1.35(7)$ &    $1.11(4)$ &   $1.39(11)$ &    $0.96(6)$ \\
\hline
\end{tabular}

\caption{Non-perturbative results of the renormalization constants
for the sextet and triplet components of the lattice energy-momentum
tensor.}   
\label{tab:zetas}
\end{table}  

In this Appendix we collect the Monte Carlo results at fixed bare lattice 
parameters of the quantities that determine the renormalization constants according to Eqs.~\eqref{eq:ZG6}--\eqref{eq:ZF3}, and of the
resulting renormalization factors.

The thermal expectation values of the relevant sextet and
triplet components of the bare lattice energy-momentum tensor at given
values of $L_0/a$ and $\beta=6/g_0^2$, and twist phase parameter 
$\theta_0=\thA= 0$, are reported in Table~\ref{tab:tensor_thA}. The results
for $\beta<15$ have been measured on lattice ensembles with spatial size
$L/a=288$, that were previously generated for an independent study~\cite{DallaBrida:2020gux}.
The results at $\beta=15$ have been obtained from dedicated Monte Carlo 
simulations of lattices with spatial size $L/a=144$.
The same observables for the twist phase parameter 
$\theta_0=\thB=3\pi/10$, are reported in Table~\ref{tab:tensor_thB}.
The latter have been computed through Monte Carlo simulations on lattices
with spatial size $L/a=144$.
The statistics for the $L/a=288$ ensembles amounts to about 100-150 
measurements at $L_0/a=4,6$, and 250 measurements for $L_0/a=8$. This 
results in a precision of about $0.2\%$ for the gluonic components (both
sextet and triplet) and of about $0.05\%$ for the fermionic ones at 
$L_0/a=4$; the same observables at $L_0/a=6$ have accuracies of about 
$0.9\%$ and $0.2\%$, respectively, and finally, at $L_0/a=8$, these figures
increase to about $1.5\%$ and $0.3\%$, respectively.
The collected statistics for the $L/a=144$ ensembles amounts to about 300 measurements on lattices with $L_0/a=4$, about 2000 measurements 
at $L_0/a=6$, and about 5000 measurements at $L_0/a=8$. At $L_0/a=4$, the 
obtained precision for the gluonic components is $0.4\%$ or so, and for the 
fermionic ones is about $0.1\%$. At $L_0/a=6$, the relative errors increase
to $0.7\%$ and $0.2\%$ respectively, and finally at $L_0/a=8$ we have 
precisions of about $1\%$ and $0.3\%$.

The values for the derivative of the free-energy density at fixed bare parameters $L_0/a$ and $\beta$, and twist phase $\theta_0=\theta_0^A$ are
collected in Table~\ref{tab:dfdv}. Results at $\beta<15$ are taken from 
Refs.~\cite{Bresciani:2025vxw,Bresciani:2025mcu}, while those at 
$\beta=15$ have been obtained from dedicated Monte Carlo simulations.
The accuracy of this quantity is about $0.05\%$ at $L_0/a=4$, $0.3\%$ at
$L_0/a=6$ and $0.5\%$ at $L_0/a=8$. 

Table~\ref{tab:dfdv} also reports the values of the integral in the twist
phase ${\cal V}_{0,1}^{AB}$ computed on lattices with shift $\vec\xi=(1,0,0)$.
The latter is computed through a 3-point 
Gauss-Legendre quadrature, as described in Subsection~\ref{ssec:NuAB}. 
The numerical results
for the discrete derivative in the shift of the thermal expectation value
of the conserved vector current $V_0$ are reported as well, at the three 
values of $\theta_0$ prescribed by the chosen quadrature rule. These 
results have been obtained from Monte Carlo simulations of lattices with 
spatial size $L/a=144$. At a given value of $\beta$, and for each value
of the twist phase and shift at which 
$\Delta\corr{V_0}_{\scriptscriptstyle\vec\xi,\theta_0}/\Delta\xi_k$ has 
to be determined,  we have collected a statistics of about 100 
measurements for $L_0/a=4$ and about 200 measurements for $L_0/a=6,8$.
The resulting relative errors for $\NuAB$ are about $0.05\%$ at $L_0/a=4$,
$0.2\%$ at $L_0/a=6$ and $0.6\%$ at $L_0/a=8$. 

Finally, the values of the renormalization constants at fixed $L_0/a$ and 
$\beta$, obtained using the results described so far in 
Eqs.~\eqref{eq:ZG6}--\eqref{eq:ZF3}, are reported in 
Table~\ref{tab:zetas}. At $L_0/a=6$, 
the relative errors on these results is about $3\%-4\%$ for the gluonic 
sextet and triplet cases, and about $2\%-3\%$ for the fermionic cases.
At $L_0/a=4$, those figures are about halved, while at $L_0/a=8$ they are
about doubled. These uncertainties are of statistical nature, and the 
largest contribution (more than $90\%$ of the variance) comes from the 
Monte Carlo fluctuations of the matrix elements of the energy-momentum 
tensor. This is related to the fact that 
Eqs.~\eqref{eq:ZG6}--\eqref{eq:ZF3} depend on differences of products of
matrix elements determined at the two twist angles $\theta_0^A$ and 
$\theta_0^B$. Therefore the signal partly cancels while the errors, which
are almost independent on the twist angle, sum in quadrature.

\section{Systematic effects from the Gaussian quadratures}
\label{app:NuAB_int}
\begin{figure*}[t]
\begin{center}
\begin{minipage}{0.5\columnwidth}
\includegraphics[width=0.95\textwidth]{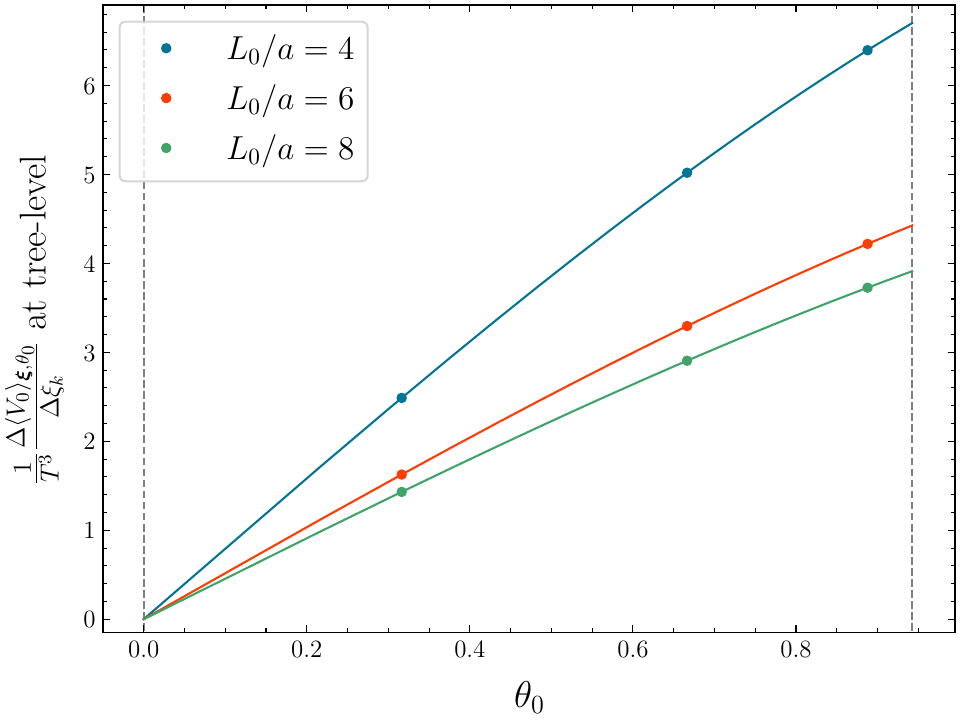}
\end{minipage}%
\begin{minipage}{0.5\columnwidth}
\includegraphics[width=0.95\textwidth]{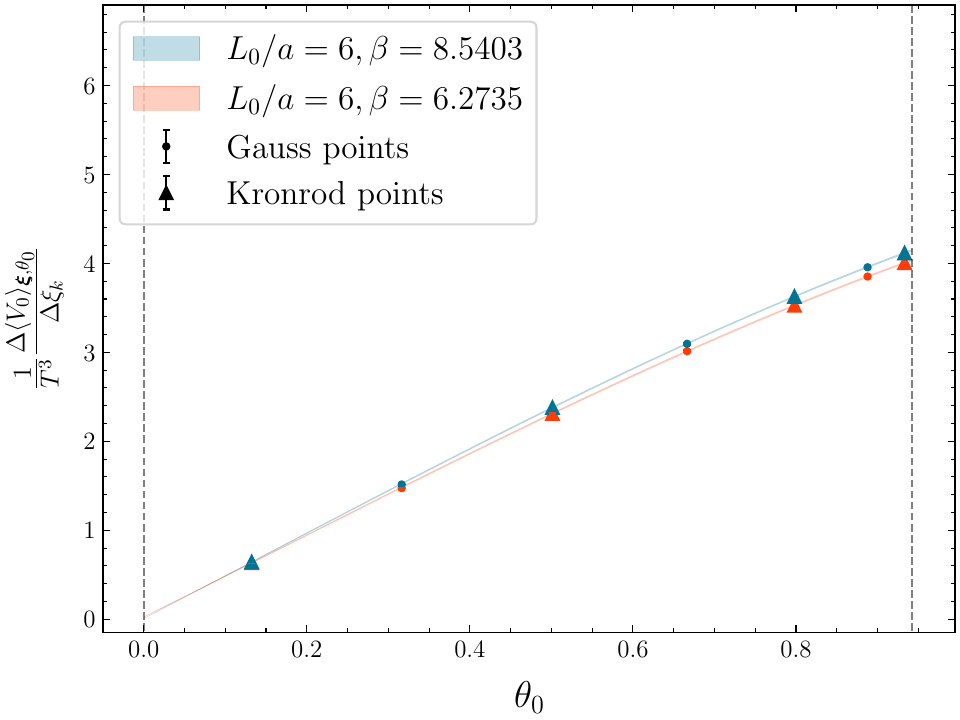}
\end{minipage}
    \caption{Left: derivative with respect to the shift of the thermal
    expectation value of the conserved vector current at tree-level in
    lattice perturbation theory. The points denote the values 
    corresponding to the nodes for the Gaussian quadrature. 
    Right: non-perturbative results at some selected bare 
    parameters. In both plots, the dashed vertical lines denote the
    integration interval.}
    \label{fig:DVcDxi}
\end{center}
\end{figure*}

In this Appendix we motivate the use of a 3-point Gauss-Legendre quadrature
for the numerical evaluation of the integral ${\cal V}_{0,k}^{AB}$
in Eq.~\eqref{V_{0,k}^{AB}num} by showing that the systematic effects due
to the approximate integration in the fermionic twist phase $\theta_0$ are
negligible with respect to the final statistical accuracy on the
renormalization constants.

The number of integration points has been set from an exploratory
study at tree-level in lattice perturbation theory. At this order, all 
terms in Eq.~\eqref{eq:dfB} can be computed analytically, and therefore the 
systematics on the numerical integration can be accessed by comparing the
two sides of that equation. To this aim, the tree-level values of the 
quantity $\Delta\corr{V_0}_{\sss\bsxi,\theta_0}/\Delta\xi_k$, appearing
in the integrand function of Eq.~\eqref{V_{0,k}^{AB}num}, for lattices with 
$L_0/a=4,6,8$ and $L/L_0=32$ are shown as a function of $\theta_0$ in the
left panel of Figure~\ref{fig:DVcDxi}. When the integration is carried out
with the chosen scheme, and the derivatives of the free-energy in the
shift appearing in Eq.~\eqref{eq:dfB} are computed at tree-level,
that equation is satisfied up to relative corrections of the order of
$10^{-10}$, thus much more precisely than needed to us.

\begin{table}[t]
\centering
\begin{tabular}{|c|c|c|c|c|c|}
    \hline
    $\beta=6/g_0^2$ & quadrature & ${\cal Z}_G^{\{6\}}$ & ${\cal Z}_F^{\{6\}}$ 
    & ${\cal Z}_G^{\{3\}}$ & ${\cal Z}_F^{\{3\}}$ \rule[-6pt]{0pt}{20pt} \\
    \hline
    \multirow{ 2}{*}{$8.5403$}
    & Gauss-Legendre & 1.51(4) &  1.129(23) &    1.35(5) &  1.053(25) \\
    & Gauss-Kronrod  & 1.51(4) &  1.131(23) &    1.35(5) &  1.055(25) \\
    \hline
    \multirow{ 2}{*}{$6.2735$}
    & Gauss-Legendre & 1.76(6) &  1.162(28) &    1.52(6) &    1.07(3) \\
    & Gauss-Kronrod  & 1.75(6) &  1.164(28) &    1.51(6) &    1.07(3) \\
    \hline
\end{tabular}

\caption{Results for the renormalization constants at $L_0/a=6$ and two
values of $\beta$, where ${\cal V}_{0,k}^{AB}$ is computed 
with either the Gauss-Legendre or the Gauss-Kronrod quadrature.}
\label{tab:ZT_gk_check}
\end{table}

At the non-perturbative level, the same quantity is confirmed to be a very
smooth function of $\theta_0$ as well, as shown in the right panel of 
Figure~\ref{fig:DVcDxi} for $L_0/a=6$, $\beta=8.5403$ and
$\beta=6.2735$. In this case, to estimate the systematics from the
integration by quadratures, we have implemented a Gauss-Kronrod quadrature
integration scheme. Indeed, while the 3-point Gauss-Legendre rule
integrates exactly polynomials up to order 5, the Gauss-Kronrod rule 
increases the order to 10 by adding 4 points to the Gauss-Legendre 
ones~\cite{Rabinowitz1980TheED}. The additional points for the 
Gauss-Kronrod rule have been generated by Monte Carlo simulations 
similarly to the Gauss-Legendre ones, see Subsection~\ref{ssec:NuAB}.
In the right panel of Figure~\ref{fig:DVcDxi} we show the
Gauss-Legendre points together with the additional Gauss-Kronrod
ones and, to facilitate the comparison, polynomial interpolations
of the latter (shadowed bands). 
We have then determined the renormalization constants
by using either integration rule for ${\cal V}_{0,k}^{AB}$.
The results are shown in Table~\ref{tab:ZT_gk_check}: the tiny 
discrepancies among the results from the two integration schemes are, in 
the worst case, 10 times smaller than the statistical accuracy. 
We have further verified that the shape of the quantity 
$\Delta\corr{V_0}_{\bsxi,\theta_0}/\Delta\xi_k$ depends very mildly 
on $\beta$ and $L_0/a$, so that a similar analysis would lead to analogous
results for all the lattices considered. These results therefore guarantee
that the chosen quadrature scheme is sufficient for the systematic
uncertainty of the numerical integration to be negligible with respect to
the statistical accuracy attained on the renormalization constants at all
values of bare parameters considered in this work.

\section{Perturbative improvement}
\label{app:appPT}

This Appendix provides the details of the perturbative computations needed 
to improve to one-loop order the definition of the renormalization constants
of the traceless components of the energy-momentum tensor. By using the 
perturbative results in Ref.~\cite{DallaBrida:2020gux}, and by setting
the shift vector to $\bsxi=(1,0,0)$, we evaluated at one-loop order for 
$L_0/a=4,6$ and $8$ both the discrete derivative of the free-energy density 
at $\theta_0^A$ and $\theta_0^B$, and the thermal expectation values 
$ \langle {T}_{01}^{X,\{i\}} \rangle_{\scriptscriptstyle \vec\xi,\theta_0}$
for $i=3,6$, $X=F,G$, and $\theta_0=\theta_0^A, \theta_0^B$.
By writing the perturbative expansions of the renormalization constants as
${\cal Z}^{\{i\}}_{X} =  {\cal Z}^{\{i\}(0)}_{X} + 
{\cal Z}^{\{i\}(1)}_{X} \, g_0^2$, the perturbative counterparts of 
Eqs.~\eqref{eq:ZG6}--\eqref{eq:ZF3} to one-loop order can be written as 
\begin{equation}
  {\cal Z}^{\{i\}(0)}_{X} =\dfrac{N^{\{i\}(0)}_X}{D^{\{i\}(0)}_X}\, , 
 \quad
  {\cal Z}^{\{i\}(1)}_{X}(L_0/a)=\dfrac{N^{\{i\}(0)}_X}{D^{\{i\}(0)}_X} 
  \left(\dfrac{N^{\{i\}(1)}_X}{N^{\{i\}(0)}_X}
  -\dfrac{D^{\{i\}(1)}_X}{D^{\{i\}(0)}_X} \right)\,,
  \label{eq:Zpert}
\end{equation}
where the expansions $N=N^{\{i\}(0)}_X + N^{\{i\}(1)}_X\, g_0^2$ and  
$D=D^{\{i\}(0)}_X + D^{\{i\}(1)}_X\, g_0^2$ of numerators and denominators
of Eqs.~\eqref{eq:ZG6}--\eqref{eq:ZF3} have been introduced, and the 
dependence on $a/L_0$ has been suppressed. Numerical results
are collected in Table~\ref{tab:Zpert} 
together with the values in the $1/L_0 \rightarrow 0$ limit obtained
in Refs.~\cite{Caracciolo:1991cp,Burgio:1996ji,Capitani:1994qn,Capitani:2002mp,Yang:2016xsb,DallaBrida:2020gux}. 

\begin{table}[t]
\centering
\begin{tabular}{|c|c|c|c|c|}
\hline
 & $L_0/a=4$  & $L_0/a=6$  & $L_0/a=8$  & $1/L_0\to0$ \\
\hline
${\cal Z}_G^{\{6\}(0)}$ \rule[-6pt]{0pt}{20pt} &    $1.1058312743$
 &    $1.0472332205$ &    $1.0301170272$ &               $1$ \\
${\cal Z}_F^{\{6\}(0)}$ \rule[-6pt]{0pt}{20pt} &    $1.5977033529$
 &    $1.1758524353$ &    $1.0700707660$ &               $1$ \\
${\cal Z}_G^{\{6\}(1)}$ \rule[-6pt]{0pt}{20pt} &      $0.40085(5)$
 &     $0.37740(10)$ &     $0.37050(20)$ &     $0.361285(8)$ \\
${\cal Z}_F^{\{6\}(1)}$ \rule[-6pt]{0pt}{20pt} &  $-0.0183950(20)$
 &  $-0.0070830(20)$ &     $0.006702(5)$ &    $0.014085(11)$ \\
\hline
${\cal Z}_G^{\{3\}(0)}$ \rule[-6pt]{0pt}{20pt} &    $1.0534476125$
 &    $1.0369229639$ &    $1.0257243477$ &               $1$ \\
${\cal Z}_F^{\{3\}(0)}$ \rule[-6pt]{0pt}{20pt} &    $1.4629519492$
 &    $1.0931360060$ &    $1.0214585873$ &               $1$ \\
${\cal Z}_G^{\{3\}(1)}$ \rule[-6pt]{0pt}{20pt} &      $0.35559(6)$
 &      $0.33314(9)$ &     $0.32560(20)$ &      $0.31465(6)$ \\
${\cal Z}_F^{\{3\}(1)}$ \rule[-6pt]{0pt}{20pt} &   $-0.056400(10)$
 &     $-0.04026(4)$ &     $-0.02943(6)$ &    $-0.02907(27)$ \\
\hline
\end{tabular}

\caption{Perturbative coefficients at tree-level, ${\cal Z}_X^{\{i\}(0)}$, 
and one-loop order, ${\cal Z}_X^{\{i\}(1)}$, of the renormalization 
constants of the fermionic and gluonic components, $X=G,F$, of the lattice 
energy-momentum tensor in the triplet and sextet representations, $i=3,6$. 
The data correspond to the thermodynamic limit for the given values of
$L_0/a$.}
\label{tab:Zpert}  
\end{table}

\subsection{Thermodynamic limit}
The computations have been carried out in coordinate space with
finite spatial dimensions. The results have then been extrapolated to the thermodynamic
limit adopting different approaches depending on the quantity.

\subsubsection{Tree-level contributions}
The tree-level gluonic contribution to the free-energy density
and the gluonic components of the energy-momentum tensor in the thermodynamic limit have been obtained by replacing
summations over momenta with continuum integrals, and then performing the
integrations over spatial momenta analytically as explained in Appendix~B of Ref.~\cite{Bresciani:2025mcu}\footnote{ 
Note that the explicit form of the tree-level gluonic contribution to the free-energy density in the thermodynamic limit
is given in Eq.~(B36) in Ref.~\cite{Bresciani:2025mcu}.}.
On the other hand, finite size effects in the fermionic contribution to the free-energy density and the
fermionic component of the energy-momentum tensor are exponentially suppressed in $LT$. Therefore,
their values in the thermodynamic limit can be estimated using a sufficiently large spatial volume, where
finite-volume effects are not detectable within machine precision. We used $L/a = 384$. The $Z^{\{i\}(0)}_{X}$ then follows from Eq.~\eqref{eq:Zpert}.

\subsubsection{One-loop contributions}
At each value of $L_0/a=4,6,8$, we have first evaluated the one-loop
contributions to the renormalization constants using Eq.~\eqref{eq:Zpert}
at finite $L/a$ ranging between 96 and 384, and then we have extrapolated
them to the thermodynamic limit using a polynomial in $a/L$ as fit 
function. The final results are reported in the columns at fixed $L_0/a$ of
Table~\ref{tab:Zpert}. The error bars have been estimated by changing the
order of the fit polynomial from 2 to 4, by considering several fit ranges 
in $a/L$, and also by considering the non-uniform Richardson--Romberg
extrapolation to remove power-like corrections in $a/L$ up to $O((a/L)^4)$. 
The quoted systematic uncertainties are tiny compared to the 
statistical errors of the Monte Carlo results for the renormalization
constants, and can be safely neglected when improving the latter according
to Eq.~\eqref{eq:ZTptimpr} with no detectable impact.

\section{Details on the $1/L_0\to0$ extrapolations}
\label{app:appExtrap}
\begin{table}[t!]
\centering
\begin{tabular}{|c|c|c|}
\hline
\texttt{id} & $L_0/a$ & fit function \\ 
 \hline
\texttt{id0} \rule[-10pt]{0pt}{28pt} & $4,6,8$ & $1 + Z_{X}^{\{i\}(1)}g_0^2 +c_{X,2}^{\{i\}}\,g_0^4 +d_{X,12}^{\{i\}}\Big(\dfrac{a}{L_0}\Big) g_0^4$ \\
\texttt{id1} \rule[-10pt]{0pt}{28pt} & $4,6,8$ & $1 + Z_{X}^{\{i\}(1)}g_0^2 +c_{X,2}^{\{i\}}\,g_0^4 +d_{X,22}^{\{i\}}\Big(\dfrac{a}{L_0}\Big)^2 g_0^4$ \\
\texttt{id2} \rule[-10pt]{0pt}{28pt} & $4,6,8$ & $1 + Z_{X}^{\{i\}(1)}g_0^2 +c_{X,2}^{\{i\}}\,g_0^4 +c_{X,3}^{\{i\}}\,g_0^6 +d_{X,22}^{\{i\}}\Big(\dfrac{a}{L_0}\Big)^2 g_0^4$ \\
\texttt{id3} \rule[-10pt]{0pt}{28pt} & $4,6,8$ & $1 + Z_{X}^{\{i\}(1)}g_0^2 +c_{X,2}^{\{i\}}\,g_0^4 +c_{X,3}^{\{i\}}\,g_0^6 +\Big(\dfrac{a}{L_0}\Big)^2\left(d_{X,22}^{\{i\}}\,g_0^4 +d_{X,23}^{\{i\}}\,g_0^6\right)$ \\
\texttt{id4} \rule[-10pt]{0pt}{28pt} & $6,8$ & $1 + Z_{X}^{\{i\}(1)}g_0^2 +c_{X,2}^{\{i\}}\,g_0^4$ \\
\texttt{id5} \rule[-10pt]{0pt}{28pt} & $6,8$ & $1 + Z_{X}^{\{i\}(1)}g_0^2 +c_{X,2}^{\{i\}}\,g_0^4 +c_{X,3}^{\{i\}}\,g_0^6$ \\
\hline
\end{tabular}
\caption{Summary of the fit procedures considered for the $1/L_0\to0$
extrapolations of the renormalization constants. Each fit procedure is 
labelled by an \texttt{id} number (first column), and may include or not 
the $L_0/a=4$ points (second column). The third column shows the fit 
functions, where $X=G,F$ for gluonic or fermionic, and $i=6,3$ for sextet
or triplet representation.}
\label{tab:fit_procedures}
\end{table}

In this Appendix we report on the most representative fit procedures that
we have performed as consistency and stability checks for the 
$1/L_0\to0$ extrapolations of the renormalization constants. 
The fit procedures are labelled by an \texttt{id} number and they differ
by the dataset considered (either $L_0/a\geq 4$ or $L_0/a>4$ points) and by 
the parameters included in the fit function, as summarized in 
Table~\ref{tab:fit_procedures}. In all cases, the constant and the 
$O(g_0^2)$ coefficients of the $1/L_0\to0$ part of the fit functions 
have been set to the known values in perturbation theory, given in 
Table~\ref{tab:Zpert}. The uncertainties quoted in that table are orders of
magnitude smaller than the statistical fluctuations of the non-perturbative
results entering the fits, and give a completely negligible contribution to
the errors of the fitted coefficients.

\subsection{The sextet case: $Z^{\{6\}}_{G}$ and $Z^{\{6\}}_{F}$}

\begin{figure}[ht!]

    \centering

    \begin{minipage}{0.5\columnwidth}
    \centering
    \includegraphics[width=0.95\textwidth]{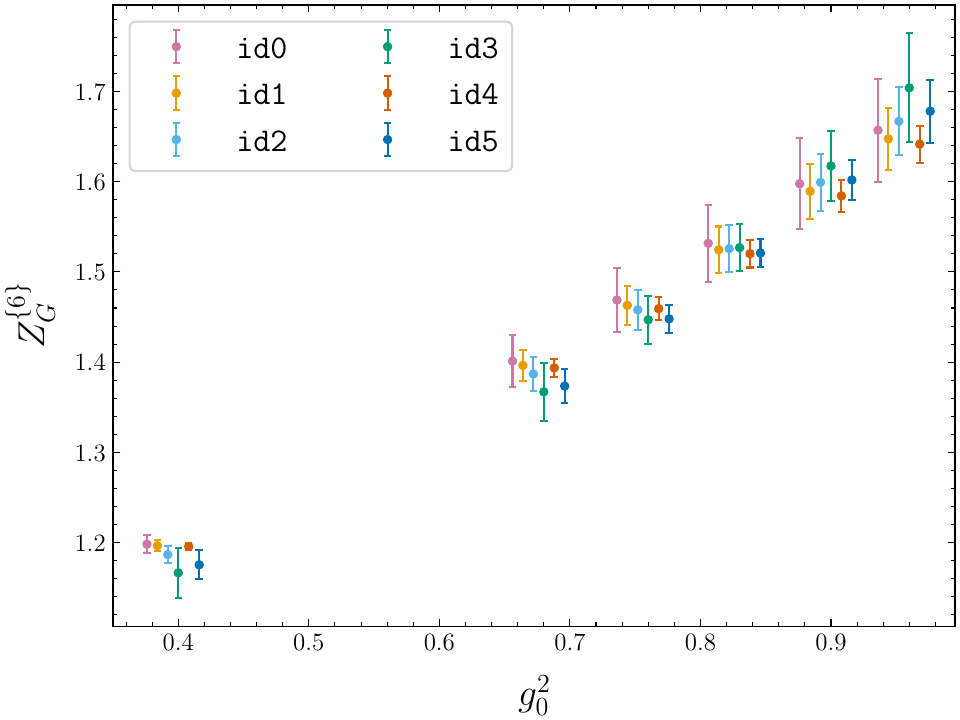}
    \end{minipage}\hfill
    \begin{minipage}{0.5\columnwidth}
    \centering
    \includegraphics[width=0.95\textwidth]{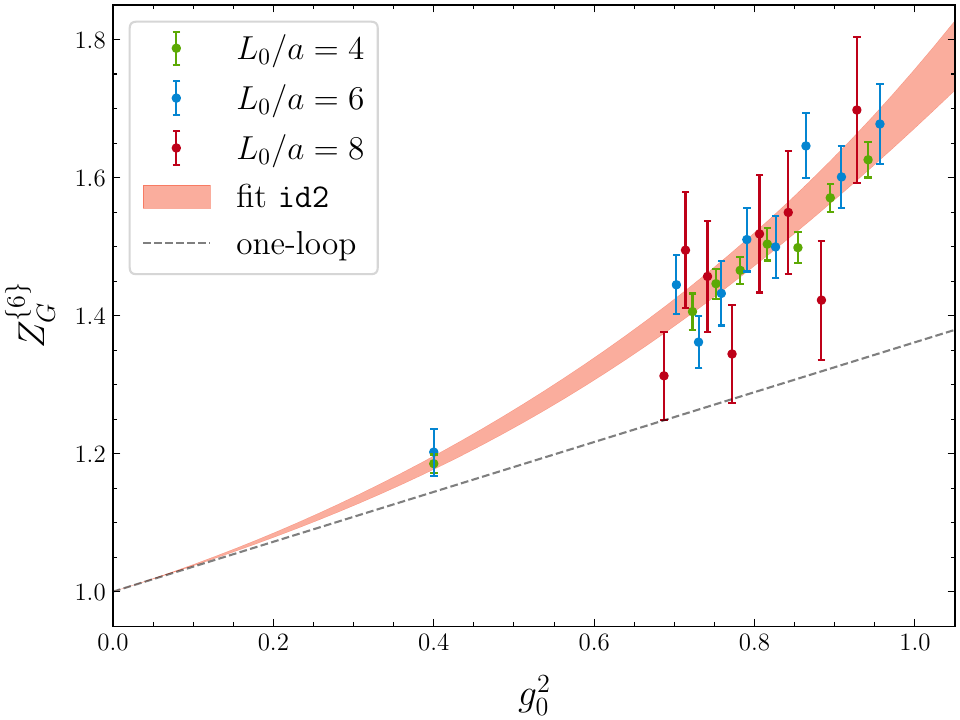}
    \end{minipage}
    \caption{Left panel: comparison of results in the $1/L_0\to0$ limit for 
    $Z_G^{\{6\}}$, obtained from several different fit strategies. For
    better readability, data have been shifted horizontally by
    $0.008\times(n-3)$ where $n$ stands for the $\texttt{id}$ number of 
    the fit.
    Right panel: the one-loop improved non-perturbative results for
    $Z_G^{\{6\}}$ at fixed $L_0/a$ are compared with the best fit
    procedure, and with perturbation theory to one-loop order.}
    \label{fig:ZG6_fit_results}

    \vspace{0.5cm}

    \begin{minipage}{\columnwidth}
    \centering
    \begin{tabular}{|c|c|c|c|c|c|c|}
\hline
                                   $g_0^2$ & \texttt{id0} & \texttt{id1} & \texttt{id2} & \texttt{id3} & \texttt{id4} & \texttt{id5} \\
\hline
                                    $0.40$ & $ 1.198(10)$ & $  1.197(6)$ & $ 1.187(10)$ & $ 1.167(28)$ & $  1.196(4)$ & $ 1.175(16)$ \\
                                    $0.68$ & $ 1.401(29)$ & $ 1.396(17)$ & $ 1.387(19)$ & $   1.37(3)$ & $ 1.394(10)$ & $ 1.374(19)$ \\
                                    $0.76$ & $   1.47(4)$ & $ 1.463(22)$ & $ 1.458(22)$ & $ 1.447(26)$ & $ 1.459(13)$ & $ 1.448(16)$ \\
                                    $0.83$ & $   1.53(4)$ & $ 1.524(26)$ & $ 1.526(26)$ & $ 1.527(26)$ & $ 1.520(15)$ & $ 1.521(15)$ \\
                                    $0.90$ & $   1.60(5)$ & $   1.59(3)$ & $   1.60(3)$ & $   1.62(4)$ & $ 1.584(18)$ & $ 1.602(22)$ \\
                                    $0.96$ & $   1.66(6)$ & $   1.65(3)$ & $   1.67(4)$ & $   1.70(6)$ & $ 1.642(20)$ & $   1.68(3)$ \\
\hline
        $c_{G,2}^{\{6\}}$ \rule[-6pt]{0pt}{20pt} & $   0.34(6)$ & $   0.33(4)$ & $  0.20(10)$ & $   -0.0(3)$ & $ 0.320(22)$ & $  0.08(19)$ \\
        $c_{G,3}^{\{6\}}$ \rule[-6pt]{0pt}{20pt} & $          $ & $          $ & $  0.15(11)$ & $    0.4(4)$ & $          $ & $  0.30(23)$ \\
       $d_{G,12}^{\{6\}}$ \rule[-6pt]{0pt}{20pt} & $ -0.14(27)$ & $          $ & $          $ & $          $ & $          $ & $          $ \\
       $d_{G,22}^{\{6\}}$ \rule[-6pt]{0pt}{20pt} & $          $ & $   -0.4(7)$ & $   -0.4(7)$ & $      4(6)$ & $          $ & $          $ \\
       $d_{G,23}^{\{6\}}$ \rule[-6pt]{0pt}{20pt} & $          $ & $          $ & $          $ & $     -6(7)$ & $          $ & $          $ \\
\hline
$\chi^2/\chi^2_{\rm exp}$ \rule[-6pt]{0pt}{20pt} & $      0.98$ & $      0.97$ & $      0.94$ & $      0.95$ & $      1.19$ & $      1.16$ \\
\hline
\end{tabular}

    \captionof{table}{
    Results for the renormalization constant $Z_G^{\{6\}}$ for 
    different $1/L_0\to0$ extrapolation procedures. Each column
    corresponds to the fit labelled by the \texttt{id}-number in the first
    row. The first six rows report the numerical values of the
    renormalization constants at six values of $g_0^2$ between $0.40$ and
    $0.96$. The remaining rows show the parameters and the quality of the
    fits.}
    \label{tab:ZG6_fit_results}
    \end{minipage}

\end{figure}

\begin{figure}[ht!]

    \centering

    \begin{minipage}{0.5\columnwidth}
    \centering
    \includegraphics[width=0.95\textwidth]{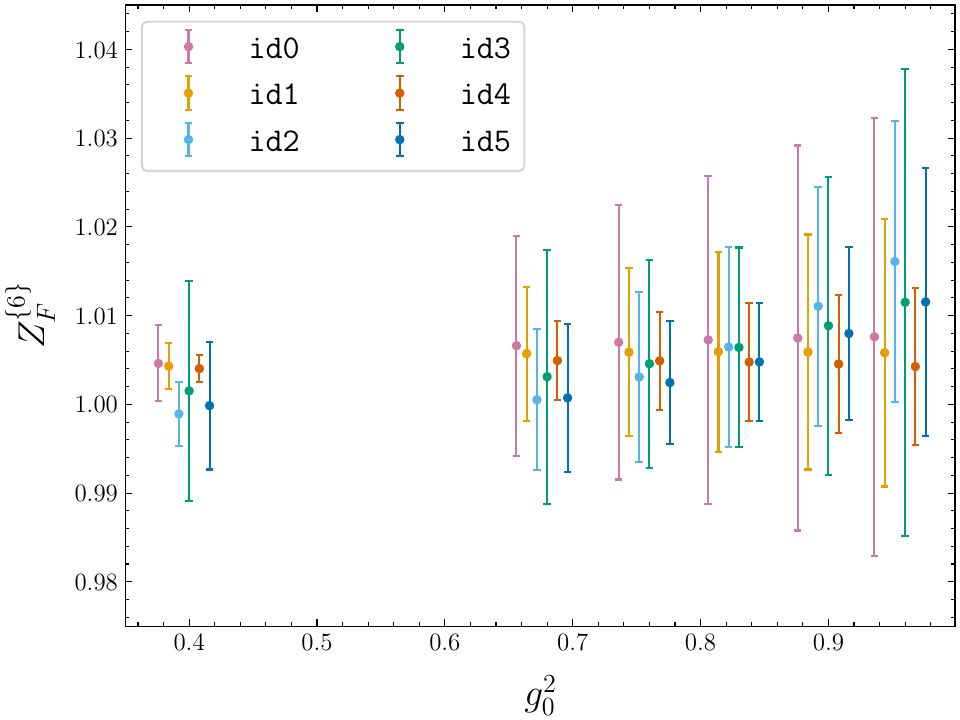}
    \end{minipage}\hfill
    \begin{minipage}{0.5\columnwidth}
    \centering
    \includegraphics[width=0.95\textwidth]{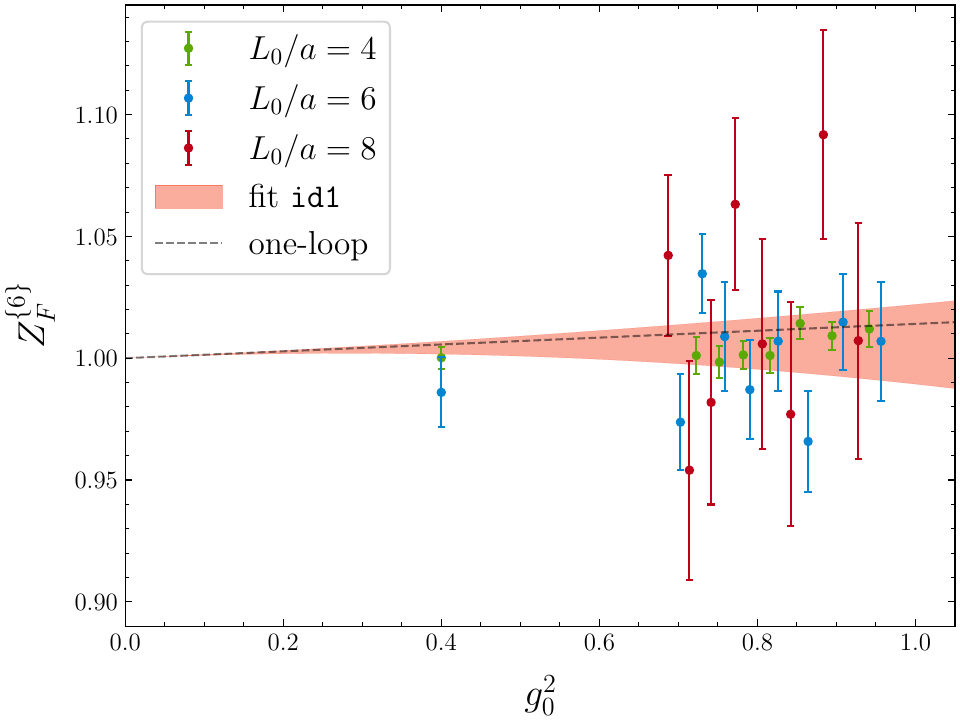}
    \end{minipage}
    \caption{Plots as in Figure~\ref{fig:ZG6_fit_results}, but for
    the renormalization constant $Z_F^{\{6\}}$.}
    \label{fig:ZF6_fit_results}

    \vspace{0.5cm}

    \begin{minipage}{\columnwidth}
    \centering
    \begin{tabular}{|c|c|c|c|c|c|c|}
\hline
                                   $g_0^2$ & \texttt{id0} & \texttt{id1} & \texttt{id2} & \texttt{id3} & \texttt{id4} & \texttt{id5} \\
\hline
                                    $0.40$ & $  1.005(4)$ & $1.0043(26)$ & $  0.999(4)$ & $ 1.002(12)$ & $1.0040(15)$ & $  1.000(7)$ \\
                                    $0.68$ & $ 1.007(12)$ & $  1.006(8)$ & $  1.001(8)$ & $ 1.003(14)$ & $  1.005(4)$ & $  1.001(8)$ \\
                                    $0.76$ & $ 1.007(15)$ & $  1.006(9)$ & $ 1.003(10)$ & $ 1.005(12)$ & $  1.005(6)$ & $  1.002(7)$ \\
                                    $0.83$ & $ 1.007(18)$ & $ 1.006(11)$ & $ 1.006(11)$ & $ 1.006(11)$ & $  1.005(7)$ & $  1.005(7)$ \\
                                    $0.90$ & $ 1.007(22)$ & $ 1.006(13)$ & $ 1.011(13)$ & $ 1.009(17)$ & $  1.005(8)$ & $ 1.008(10)$ \\
                                    $0.96$ & $ 1.008(25)$ & $ 1.006(15)$ & $ 1.016(16)$ & $ 1.011(26)$ & $  1.004(9)$ & $ 1.012(15)$ \\
\hline
        $c_{F,2}^{\{6\}}$ \rule[-6pt]{0pt}{20pt} & $-0.006(27)$ & $-0.008(16)$ & $  -0.07(3)$ & $ -0.04(15)$ & $-0.010(10)$ & $  -0.06(9)$ \\
        $c_{F,3}^{\{6\}}$ \rule[-6pt]{0pt}{20pt} & $          $ & $          $ & $   0.08(4)$ & $  0.04(18)$ & $          $ & $  0.06(10)$ \\
       $d_{F,12}^{\{6\}}$ \rule[-6pt]{0pt}{20pt} & $ -0.01(11)$ & $          $ & $          $ & $          $ & $          $ & $          $ \\
       $d_{F,22}^{\{6\}}$ \rule[-6pt]{0pt}{20pt} & $          $ & $ -0.00(28)$ & $ -0.03(28)$ & $  -0.6(25)$ & $          $ & $          $ \\
       $d_{F,23}^{\{6\}}$ \rule[-6pt]{0pt}{20pt} & $          $ & $          $ & $          $ & $      1(3)$ & $          $ & $          $ \\
\hline
$\chi^2/\chi^2_{\rm exp}$ \rule[-6pt]{0pt}{20pt} & $      1.23$ & $      1.23$ & $      1.08$ & $      1.13$ & $      1.38$ & $      1.45$ \\
\hline
\end{tabular}

    \captionof{table}{Data as in Table~\ref{tab:ZG6_fit_results}, but for
    the renormalization constant $Z_F^{\{6\}}$.}
    \label{tab:ZF6_fit_results}
    \end{minipage}

\end{figure}

We start from the sextet gluonic case $Z_{G}^{\{6\}}$.
As discussed in Section~\ref{sec:theory}, in the extrapolation at
fixed $g_0^2$ a linear term in $a/L_0$ cannot be excluded in principle.
Therefore, a simple choice for the fit is a linear one in $a/L_0$ with two
free parameters at $O(g_0^4)$, one for the parametrization in the 
$1/L_0\to0$ limit (on top of the known one-loop order coefficients),
denoted by $c_{G,2}^{\{6\}}$, and one accounting for the dependence of the
$O(a/L_0)$ discretization effects on the bare coupling, called 
$d_{G,12}^{\{6\}}$. This fit is labelled as \texttt{id0}. As reported in
Table~\ref{tab:ZG6_fit_results}, the resulting fit quality is very good and
the relative error on the renormalization constants ranges from about
$0.8\%$ to $3.6\%$ in the simulated interval $0.40\leq g_0^2\leq 0.96$.

The relative relevance of any $O(a/L_0)$ terms with respect to the 
$O((a/L_0)^2)$ ones is suppressed by the factor $a\Lambda_{\rm QCD}$,
which drops from $O(0.01)$ for $g_0^2\sim0.9$ to $O(0.001)$ for 
$g_0^2\sim0.7$ and even less at smaller values of bare coupling squared.
We have thus extrapolated our points by considering a quadratic fit too
(fit \texttt{id1}), and the result is as good as \texttt{id0} in terms of 
fit quality, with central values of the renormalization constant perfectly
compatible and errors about $40\%$ smaller (see the respective columns of
Table~\ref{tab:ZG6_fit_results}). The results from the two fits for the
parameter $c_{G,2}^{\{6\}}$ are in perfect agreement, while the 
discretization effects are compatible with zero in both cases. Given these
results, for our final fit procedure we have considered a quadratic term.

In order to check the stability of the fit with respect to the choice of 
the parametrization in the bare coupling, we have added a term of 
$O(g_0^6)$, with coefficient $c_{G,3}^{\{6\}}$, to the $1/L_0\to0$
part of the fit function. This fit is labelled \texttt{id2}, its quality 
is as good as the previous ones, and the results for $Z_G^{\{6\}}$ are
perfectly compatible in terms of central values, with error magnitude
intermediate between fits \texttt{id0} and \texttt{id1}.
While a term of $O(g_0^6)$ could be present in the discretization effects
as well, the fit \texttt{id3} shows that it cannot be resolved within
the accuracy of the non-perturbative data, though the resulting values for
$Z_G^{\{6\}}$ are well compatible with the ones from the previous fit
procedures.

Finally, in order to check that the fit is not dominated by the coarsest 
non-perturbative points (which are the most precise in terms of statistical 
errors), we have discarded them and we have extrapolated $L_0/a=6,8$ data
with no parameters for the discretization effects, and one term at 
$O(g_0^4)$ in fit \texttt{id4} supplemented by a term of $O(g_0^6)$ in 
fit \texttt{id5}. The results are in very good agreement with the other fit 
procedures, with \texttt{id4} leading to the smallest errors (about a half 
with respect to fit \texttt{id2}) while the ones of fit \texttt{id5} are 
comparable to fit \texttt{id2}. 

The results of $Z_G^{\{6\}}$ from the fits at the selected values of
$g_0^2$ of Table~\ref{tab:ZG6_fit_results} are represented in the left
panel of Figure~\ref{fig:ZG6_fit_results}. Given the excellent agreement 
of all the different fit procedures, we have selected the fit \texttt{id2}
as the final best fit. The latter is represented as the red band in the 
right panel of Figure~\ref{fig:ZG6_fit_results}, in comparison to the 
perturbative prediction and to the non-perturbative  data at fixed $L_0/a$ 
entering the fit procedure, after the subtraction of discretization 
effects to one-loop order described in 
Subsection~\ref{ssec:Perturbative improvement}.

We have carried out the same analysis for the renormalization constant
$Z_F^{\{6\}}$, and analogous conclusions hold in that case too. The 
results for the different fits are given in Table~\ref{tab:ZF6_fit_results}
and are represented in the left panel of Figure~\ref{fig:ZF6_fit_results}
at some selected values of the bare coupling squared. Given the agreement 
with the perturbative result to one-loop order, in this case one parameter
only at $O(g_0^4)$ is enough and thus we have selected fit \texttt{id1} as
the best one. The related coefficient at $O(g_0^4)$ is actually zero within 
errors and estimates the uncertainty within which the non-perturbative
results are compatible to perturbation theory. The fit is represented in 
the right panel of Figure~\ref{fig:ZF6_fit_results}, in comparison to the 
curve in perturbation theory and to the non-perturbative results at fixed
$L_0/a$, perturbatively improved to one-loop order, entering the fit
procedure.

\begin{figure}[ht!]

    \centering

    \begin{minipage}{0.5\columnwidth}
    \centering
    \includegraphics[width=0.95\textwidth]{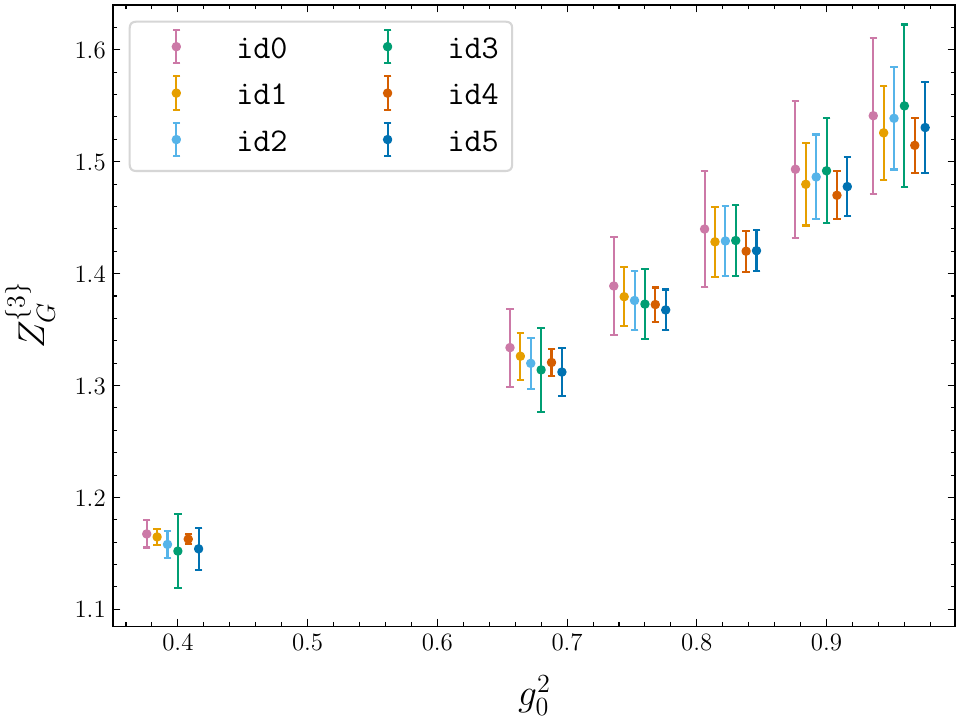}
    \end{minipage}\hfill
    \begin{minipage}{0.5\columnwidth}
    \centering
    \includegraphics[width=0.95\textwidth]{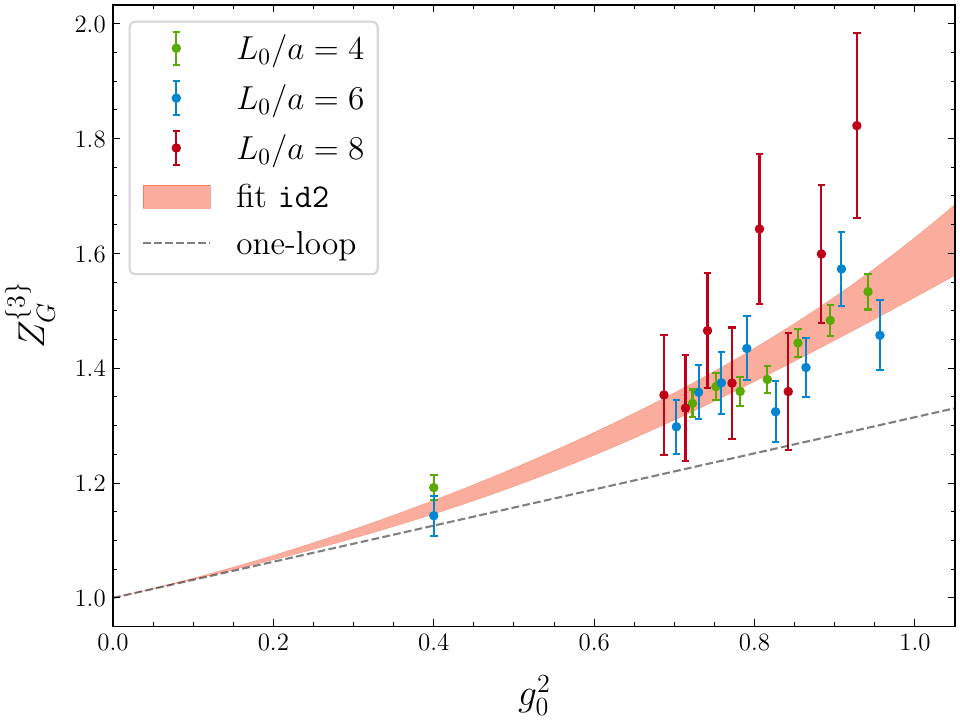}
    \end{minipage}
    \caption{Plots as in Figure~\ref{fig:ZG6_fit_results}, but for
    the renormalization constant $Z_G^{\{3\}}$.}
    \label{fig:ZG3_fit_results}

    \vspace{0.5cm}

    \begin{minipage}{\columnwidth}
    \centering
    \begin{tabular}{|c|c|c|c|c|c|c|}
\hline
                                   $g_0^2$ & \texttt{id0} & \texttt{id1} & \texttt{id2} & \texttt{id3} & \texttt{id4} & \texttt{id5} \\
\hline
                                    $0.40$ & $ 1.167(12)$ & $  1.165(7)$ & $ 1.158(12)$ & $   1.15(3)$ & $  1.163(4)$ & $ 1.154(19)$ \\
                                    $0.68$ & $   1.33(3)$ & $ 1.326(21)$ & $ 1.320(23)$ & $   1.31(4)$ & $ 1.321(12)$ & $ 1.312(22)$ \\
                                    $0.76$ & $   1.39(4)$ & $ 1.379(26)$ & $ 1.376(27)$ & $   1.37(3)$ & $ 1.372(15)$ & $ 1.368(18)$ \\
                                    $0.83$ & $   1.44(5)$ & $   1.43(3)$ & $   1.43(3)$ & $   1.43(3)$ & $ 1.420(18)$ & $ 1.420(18)$ \\
                                    $0.90$ & $   1.49(6)$ & $   1.48(4)$ & $   1.49(4)$ & $   1.49(5)$ & $ 1.470(21)$ & $ 1.478(27)$ \\
                                    $0.96$ & $   1.54(7)$ & $   1.53(4)$ & $   1.54(5)$ & $   1.55(7)$ & $ 1.515(24)$ & $   1.53(4)$ \\
\hline
        $c_{G,2}^{\{3\}}$ \rule[-6pt]{0pt}{20pt} & $   0.26(8)$ & $   0.24(5)$ & $  0.16(13)$ & $    0.1(4)$ & $ 0.231(26)$ & $  0.13(22)$ \\
        $c_{G,3}^{\{3\}}$ \rule[-6pt]{0pt}{20pt} & $          $ & $          $ & $  0.10(14)$ & $    0.2(5)$ & $          $ & $  0.13(27)$ \\
       $d_{G,12}^{\{3\}}$ \rule[-6pt]{0pt}{20pt} & $   -0.1(3)$ & $          $ & $          $ & $          $ & $          $ & $          $ \\
       $d_{G,22}^{\{3\}}$ \rule[-6pt]{0pt}{20pt} & $          $ & $   -0.2(8)$ & $   -0.3(8)$ & $      1(7)$ & $          $ & $          $ \\
       $d_{G,23}^{\{3\}}$ \rule[-6pt]{0pt}{20pt} & $          $ & $          $ & $          $ & $     -2(9)$ & $          $ & $          $ \\
\hline
$\chi^2/\chi^2_{\rm exp}$ \rule[-6pt]{0pt}{20pt} & $      1.10$ & $      1.10$ & $      1.13$ & $      1.18$ & $      1.19$ & $      1.25$ \\
\hline
\end{tabular}

    \captionof{table}{Data as in Table~\ref{tab:ZG6_fit_results}, but for
    the renormalization constant $Z_G^{\{3\}}$.}
    \label{tab:ZG3_fit_results}
    \end{minipage}

\end{figure}

\begin{figure}[ht!]

    \centering

    \begin{minipage}{0.5\columnwidth}
    \centering
    \includegraphics[width=0.95\textwidth]{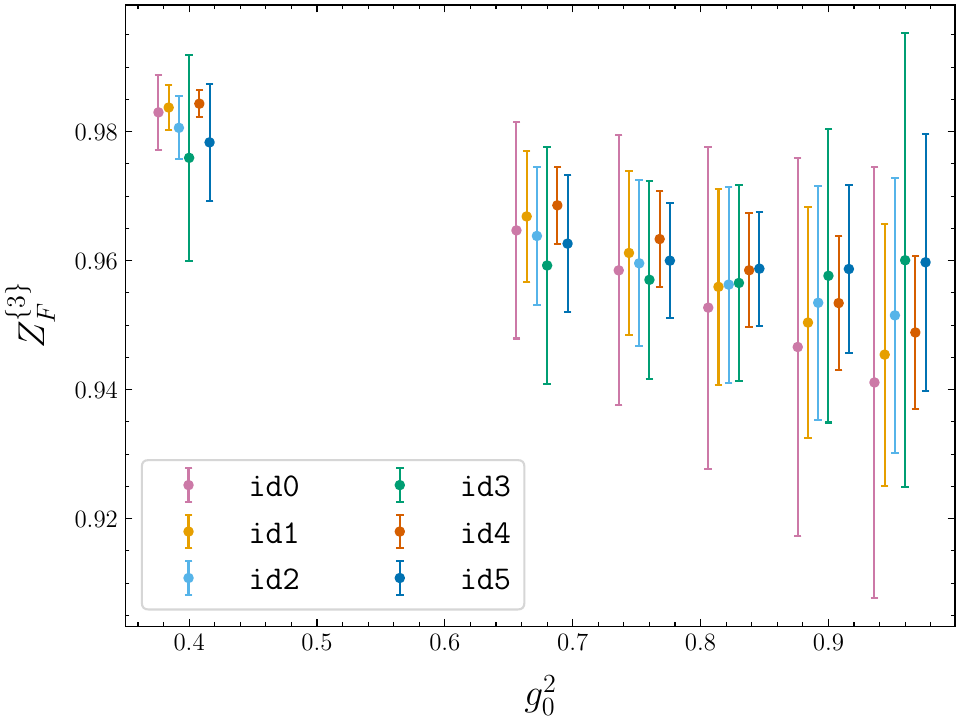}
    \end{minipage}\hfill
    \begin{minipage}{0.5\columnwidth}
    \centering
    \includegraphics[width=0.95\textwidth]{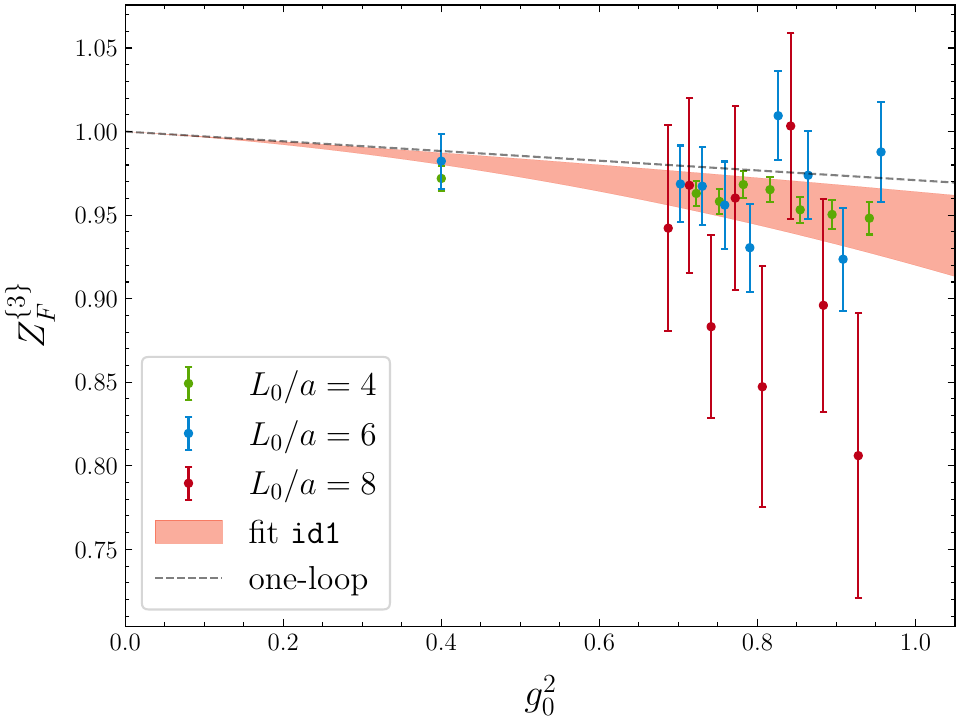}
    \end{minipage}
    \caption{Plots as in Figure~\ref{fig:ZG6_fit_results}, but for
    the renormalization constant $Z_F^{\{3\}}$.}
    \label{fig:ZF3_fit_results}

    \vspace{0.5cm}

    \begin{minipage}{\columnwidth}
    \centering
    \begin{tabular}{|c|c|c|c|c|c|c|}
\hline
                                   $g_0^2$ & \texttt{id0} & \texttt{id1} & \texttt{id2} & \texttt{id3} & \texttt{id4} & \texttt{id5} \\
\hline
                                    $0.40$ & $  0.983(6)$ & $  0.984(4)$ & $  0.981(5)$ & $ 0.976(16)$ & $0.9843(21)$ & $  0.978(9)$ \\
                                    $0.68$ & $ 0.965(17)$ & $ 0.967(10)$ & $ 0.964(11)$ & $ 0.959(18)$ & $  0.969(6)$ & $ 0.963(11)$ \\
                                    $0.76$ & $ 0.958(21)$ & $ 0.961(13)$ & $ 0.960(13)$ & $ 0.957(15)$ & $  0.963(7)$ & $  0.960(9)$ \\
                                    $0.83$ & $ 0.953(25)$ & $ 0.956(15)$ & $ 0.956(15)$ & $ 0.957(15)$ & $  0.959(9)$ & $  0.959(9)$ \\
                                    $0.90$ & $ 0.947(29)$ & $ 0.950(18)$ & $ 0.953(18)$ & $ 0.958(23)$ & $ 0.953(10)$ & $ 0.959(13)$ \\
                                    $0.96$ & $   0.94(3)$ & $ 0.945(20)$ & $ 0.952(21)$ & $   0.96(4)$ & $ 0.949(12)$ & $ 0.960(20)$ \\
\hline
        $c_{F,2}^{\{3\}}$ \rule[-6pt]{0pt}{20pt} & $  -0.03(4)$ & $-0.029(22)$ & $  -0.07(5)$ & $ -0.12(19)$ & $-0.025(13)$ & $ -0.10(11)$ \\
        $c_{F,3}^{\{3\}}$ \rule[-6pt]{0pt}{20pt} & $          $ & $          $ & $   0.05(5)$ & $  0.12(23)$ & $          $ & $  0.09(13)$ \\
       $d_{F,12}^{\{3\}}$ \rule[-6pt]{0pt}{20pt} & $  0.03(15)$ & $          $ & $          $ & $          $ & $          $ & $          $ \\
       $d_{F,22}^{\{3\}}$ \rule[-6pt]{0pt}{20pt} & $          $ & $    0.0(4)$ & $    0.0(4)$ & $      1(3)$ & $          $ & $          $ \\
       $d_{F,23}^{\{3\}}$ \rule[-6pt]{0pt}{20pt} & $          $ & $          $ & $          $ & $     -1(4)$ & $          $ & $          $ \\
\hline
$\chi^2/\chi^2_{\rm exp}$ \rule[-6pt]{0pt}{20pt} & $      0.98$ & $      0.98$ & $      0.98$ & $      1.02$ & $      1.09$ & $      1.13$ \\
\hline
\end{tabular}

    \captionof{table}{Data as in Table~\ref{tab:ZG6_fit_results}, but for
    the renormalization constant $Z_F^{\{3\}}$.}
    \label{tab:ZF3_fit_results}
    \end{minipage}

\end{figure}

\subsection{The triplet case: $Z^{\{3\}}_{G}$ and $Z^{\{3\}}_{F}$}

For the renormalization constants of the triplet representation of the 
energy-momentum tensor, $Z_G^{\{3\}}$ and $Z_F^{\{3\}}$, the study of the
$1/L_0\to0$ extrapolation has been carried out in full similarity to 
the sextet case, and leads to analogous conclusions. The results for the
gluonic renormalization constants are 
given in Table~\ref{tab:ZG3_fit_results}, and a graphical comparison of the
different fit procedures can be found in the left panel of
Figure~\ref{fig:ZG3_fit_results}. All the extrapolations are perfectly
compatible and fit \texttt{id2} has been chosen as the final fit procedure. 
The latter is represented in the right panel of 
Figure~\ref{fig:ZG3_fit_results}, in comparison to the perturbative 
curve and to the non-perturbative results at fixed $L_0/a$, after the 
subtraction of discretization effects to one-loop order, entering the fit.

Concerning $Z_F^{\{3\}}$, the results are collected in 
Table~\ref{tab:ZF3_fit_results} and represented in the two panels of 
Figure~\ref{fig:ZF3_fit_results}. Given the closeness of the 
non-perturbative results to the perturbative curve to one-loop order, we
have selected the simpler fit procedure \texttt{id1} as the final best fit.

\subsection{Fit parameters and covariance}
We provide in Table~\ref{tab:ZT_coeff} the fit parameters that determine
the non-perturbative expressions of the renormalization constants, with 
an extended number of digits. Throughout this work, the corresponding
uncertainties and correlations have been properly computed using the tools
of Refs.~\cite{Ramos:2018vgu,Joswig:2022qfe}. To allow independent 
reproduction of our results, we nevertheless provide in 
Table~\ref{tab:ZT_cov} the relevant components of the covariance matrix.
The latter has been regularized to be positive-semidefinite by setting to 
zero its only negative eigenvalue, which is also the smallest in magnitude. 
This procedure is guaranteed to produce the closest matrix in terms of the
Frobenius norm~\cite{Cheng:1998}. We have explicitly checked that any 
effect on the uncertainties of the renormalization constants, introduced by 
this procedure, is negligible within the error of the error.
\begin{table}[t]
\centering
\begin{tabular}{|c|c||c|c|}
\hline
$c$ & value & $c$ & value \\
\hline
$c_{G,2}^{\{6\}}$ & $ 2.04778\times 10^{{-1}}$ 
& $c_{G,2}^{\{3\}}$ & $ 1.61184\times 10^{{-1}}$ \rule[-6pt]{0pt}{20pt} \\
$c_{G,3}^{\{6\}}$ & $ 1.48381\times 10^{{-1}}$ 
& $c_{G,3}^{\{3\}}$ & $ 9.96055\times 10^{{-2}}$ \rule[-6pt]{0pt}{20pt} \\
$c_{F,2}^{\{6\}}$ & $-8.37555\times 10^{{-3}}$ 
& $c_{F,2}^{\{3\}}$ & $-2.89278\times 10^{{-2}}$ \rule[-6pt]{0pt}{20pt} \\
\hline
\end{tabular}

\caption{Fit parameters entering the non-perturbative expressions of 
the renormalization constants.}
\label{tab:ZT_coeff}
\end{table}
\begin{table}[t]
\centering
\begin{tabular}{|cc|c||cc|c|}
\hline
$c_1$ & $c_2$ & cov$(c_1,c_2)$ & $c_1$ & $c_2$ & cov$(c_1,c_2)$ \\
\hline
$c_{G,2}^{\{6\}}$ & $c_{G,2}^{\{6\}}$ & $9.98759\times 10^{{-3}}$ & 
$c_{G,2}^{\{6\}}$ & $c_{G,3}^{\{6\}}$ & $-1.04726\times 10^{{-2}}$
\rule[-6pt]{0pt}{20pt} \\
$c_{G,2}^{\{6\}}$ & $c_{F,2}^{\{6\}}$ & $-5.57328\times 10^{{-4}}$ & 
$c_{G,2}^{\{6\}}$ & $c_{G,2}^{\{3\}}$ & $2.17443\times 10^{{-3}}$
\rule[-6pt]{0pt}{20pt} \\
$c_{G,2}^{\{6\}}$ & $c_{G,3}^{\{3\}}$ & $-2.39844\times 10^{{-3}}$ & 
$c_{G,2}^{\{6\}}$ & $c_{F,2}^{\{3\}}$ & $-8.06355\times 10^{{-5}}$
\rule[-6pt]{0pt}{20pt} \\
$c_{G,3}^{\{6\}}$ & $c_{G,3}^{\{6\}}$ & $1.27946\times 10^{{-2}}$ & 
$c_{G,3}^{\{6\}}$ & $c_{F,2}^{\{6\}}$ & $-2.95025\times 10^{{-6}}$
\rule[-6pt]{0pt}{20pt} \\
$c_{G,3}^{\{6\}}$ & $c_{G,2}^{\{3\}}$ & $-1.91321\times 10^{{-3}}$ & 
$c_{G,3}^{\{6\}}$ & $c_{G,3}^{\{3\}}$ & $2.63651\times 10^{{-3}}$
\rule[-6pt]{0pt}{20pt} \\
$c_{G,3}^{\{6\}}$ & $c_{F,2}^{\{3\}}$ & $3.76403\times 10^{{-5}}$ & 
$c_{F,2}^{\{6\}}$ & $c_{F,2}^{\{6\}}$ & $2.78458\times 10^{{-4}}$
\rule[-6pt]{0pt}{20pt} \\
$c_{F,2}^{\{6\}}$ & $c_{G,2}^{\{3\}}$ & $-3.73268\times 10^{{-5}}$ & 
$c_{F,2}^{\{6\}}$ & $c_{G,3}^{\{3\}}$ & $5.15275\times 10^{{-5}}$
\rule[-6pt]{0pt}{20pt} \\
$c_{F,2}^{\{6\}}$ & $c_{F,2}^{\{3\}}$ & $-1.45875\times 10^{{-5}}$ & 
$c_{G,2}^{\{3\}}$ & $c_{G,2}^{\{3\}}$ & $1.57433\times 10^{{-2}}$
\rule[-6pt]{0pt}{20pt} \\
$c_{G,2}^{\{3\}}$ & $c_{G,3}^{\{3\}}$ & $-1.66811\times 10^{{-2}}$ & 
$c_{G,2}^{\{3\}}$ & $c_{F,2}^{\{3\}}$ & $-9.42650\times 10^{{-4}}$
\rule[-6pt]{0pt}{20pt} \\
$c_{G,3}^{\{3\}}$ & $c_{G,3}^{\{3\}}$ & $2.03518\times 10^{{-2}}$ & 
$c_{G,3}^{\{3\}}$ & $c_{F,2}^{\{3\}}$ & $-1.20759\times 10^{{-6}}$
\rule[-6pt]{0pt}{20pt} \\
$c_{F,2}^{\{3\}}$ & $c_{F,2}^{\{3\}}$ & $4.95326\times 10^{{-4}}$ & 
 & & 
\rule[-6pt]{0pt}{20pt} \\
\hline
\end{tabular}
\caption{Entries of the covariance matrix among the parameters of
Table~\ref{tab:ZT_coeff}.}
\label{tab:ZT_cov}
\end{table}

\clearpage

\bibliographystyle{JHEP}
\bibliography{main.bbl}

\providecommand{\href}[2]{#2}\begingroup\raggedright\begin{thebibliography}{10}

\bibitem{Weinberg:1972lg}
S.~Weinberg, \emph{{Gravitation and cosmology: principles and applications of
  the general theory of relativity}}, John Wiley \& Sons (1972).

\bibitem{Polyakov:2018zvc}
M.V.~Polyakov and P.~Schweitzer, \emph{{Forces inside hadrons: pressure,
  surface tension, mechanical radius, and all that}},
  \href{https://doi.org/10.1142/S0217751X18300259}{\emph{Int. J. Mod. Phys. A}
  {\bfseries 33} (2018) 1830025}
  [\href{https://arxiv.org/abs/1805.06596}{{\ttfamily 1805.06596}}].

\bibitem{Landau-hydro}
L.~Landau and E.~Lifshitz, \emph{{Course of Theoretical Physics, Vol. 6: Fluid
  Mechanics}} (1987).

\bibitem{Kubo:1957mj}
R.~Kubo, \emph{{Statistical mechanical theory of irreversible processes. 1.
  General theory and simple applications in magnetic and conduction problems}},
  \href{https://doi.org/10.1143/JPSJ.12.570}{\emph{J. Phys. Soc. Jap.}
  {\bfseries 12} (1957) 570}.

\bibitem{Kubo:1957wcy}
R.~Kubo, M.~Yokota and S.~Nakajima, \emph{{Statistical-Mechanical Theory of
  Irreversible Processes. II. Response to Thermal Disturbance}},
  \href{https://doi.org/10.1143/jpsj.12.1203}{\emph{J. Phys. Soc. Jap.}
  {\bfseries 12} (1957) 1203}.

\bibitem{Caracciolo:1989pt}
S.~Caracciolo, G.~Curci, P.~Menotti and A.~Pelissetto, \emph{{The Energy
  Momentum Tensor for Lattice Gauge Theories}},
  \href{https://doi.org/10.1016/0003-4916(90)90203-Z}{\emph{Annals Phys.}
  {\bfseries 197} (1990) 119}.

\bibitem{Caracciolo:1989bu}
S.~Caracciolo, G.~Curci, P.~Menotti and A.~Pelissetto, \emph{{Renormalization
  of the Energy Momentum Tensor and the Trace Anomaly in Lattice {QED}}},
  \href{https://doi.org/10.1016/0370-2693(89)91562-1}{\emph{Phys. Lett.}
  {\bfseries B228} (1989) 375}.

\bibitem{Giusti:2015daa}
L.~Giusti and M.~Pepe, \emph{{Energy-momentum tensor on the lattice:
  Nonperturbative renormalization in Yang-Mills theory}},
  \href{https://doi.org/10.1103/PhysRevD.91.114504}{\emph{Phys. Rev. D}
  {\bfseries 91} (2015) 114504}
  [\href{https://arxiv.org/abs/1503.07042}{{\ttfamily 1503.07042}}].

\bibitem{DallaBrida:2020gux}
M.~Dalla~Brida, L.~Giusti and M.~Pepe, \emph{{Non-perturbative definition of
  the QCD energy-momentum tensor on the lattice}},
  \href{https://doi.org/10.1007/JHEP04(2020)043}{\emph{JHEP} {\bfseries 04}
  (2020) 043} [\href{https://arxiv.org/abs/2002.06897}{{\ttfamily
  2002.06897}}].

\bibitem{Giusti:2010bb}
L.~Giusti and H.B.~Meyer, \emph{{Thermal momentum distribution from path
  integrals with shifted boundary conditions}},
  \href{https://doi.org/10.1103/PhysRevLett.106.131601}{\emph{Phys. Rev. Lett.}
  {\bfseries 106} (2011) 131601}
  [\href{https://arxiv.org/abs/1011.2727}{{\ttfamily 1011.2727}}].

\bibitem{Giusti:2011kt}
L.~Giusti and H.B.~Meyer, \emph{{Thermodynamic potentials from shifted boundary
  conditions: the scalar-field theory case}},
  \href{https://doi.org/10.1007/JHEP11(2011)087}{\emph{JHEP} {\bfseries 11}
  (2011) 087} [\href{https://arxiv.org/abs/1110.3136}{{\ttfamily 1110.3136}}].

\bibitem{Giusti:2012yj}
L.~Giusti and H.B.~Meyer, \emph{{Implications of Poincare symmetry for thermal
  field theories in finite-volume}},
  \href{https://doi.org/10.1007/JHEP01(2013)140}{\emph{JHEP} {\bfseries 01}
  (2013) 140} [\href{https://arxiv.org/abs/1211.6669}{{\ttfamily 1211.6669}}].

\bibitem{Luscher:2010iy}
M.~L{\"u}scher, \emph{{Properties and uses of the Wilson flow in lattice QCD}},
  \href{https://doi.org/10.1007/JHEP08(2010)071,
  10.1007/JHEP03(2014)092}{\emph{JHEP} {\bfseries 08} (2010) 071}
  [\href{https://arxiv.org/abs/1006.4518}{{\ttfamily 1006.4518}}].

\bibitem{Suzuki:2013gza}
H.~Suzuki, \emph{{Energy{\textendash}momentum tensor from the
  Yang{\textendash}Mills gradient flow}},
  \href{https://doi.org/10.1093/ptep/ptt059}{\emph{PTEP} {\bfseries 2013}
  (2013) 083B03} [\href{https://arxiv.org/abs/1304.0533}{{\ttfamily
  1304.0533}}].

\bibitem{DelDebbio:2013zaa}
L.~Del~Debbio, A.~Patella and A.~Rago, \emph{{Space-time symmetries and the
  Yang-Mills gradient flow}},
  \href{https://doi.org/10.1007/JHEP11(2013)212}{\emph{JHEP} {\bfseries 1311}
  (2013) 212} [\href{https://arxiv.org/abs/1306.1173}{{\ttfamily 1306.1173}}].

\bibitem{Makino:2014taa}
H.~Makino and H.~Suzuki, \emph{{Lattice energy-momentum tensor from the
  Yang–Mills gradient flow—inclusion of fermion fields}},
  \href{https://doi.org/10.1093/ptep/ptu070, 10.1093/ptep/ptv095}{\emph{PTEP}
  {\bfseries 2014} (2014) 063B02}
  [\href{https://arxiv.org/abs/1403.4772}{{\ttfamily 1403.4772}}].

\bibitem{Capponi:2015ahp}
F.~Capponi, L.~Del~Debbio, A.~Patella and A.~Rago, \emph{{Renormalization
  constants of the lattice energy momentum tensor using the gradient flow}},
  \href{https://doi.org/10.22323/1.251.0302}{\emph{PoS} {\bfseries LATTICE2015}
  (2016) 302} [\href{https://arxiv.org/abs/1512.04374}{{\ttfamily
  1512.04374}}].

\bibitem{Taniguchi:2016ofw}
{\scshape WHOT-QCD} collaboration, \emph{{Exploring $N_{f}$ = 2+1 QCD
  thermodynamics from the gradient flow}},
  \href{https://doi.org/10.1103/PhysRevD.96.014509}{\emph{Phys. Rev. D}
  {\bfseries 96} (2017) 014509}
  [\href{https://arxiv.org/abs/1609.01417}{{\ttfamily 1609.01417}}].

\bibitem{Harlander:2018zpi}
R.V.~Harlander, Y.~Kluth and F.~Lange, \emph{{The two-loop energy–momentum
  tensor within the gradient-flow formalism}},
  \href{https://doi.org/10.1140/epjc/s10052-018-6415-7}{\emph{Eur. Phys. J.}
  {\bfseries C78} (2018) 944}
  [\href{https://arxiv.org/abs/1808.09837}{{\ttfamily 1808.09837}}].

\bibitem{PavanPavan:2025hzr}
P.~Pavan, O.~Kaczmarek, G.D.~Moore and C.~Schmidt, \emph{{Shear viscosity from
  quenched to full lattice QCD}},
  \href{https://doi.org/10.22323/1.466.0199}{\emph{PoS} {\bfseries LATTICE2024}
  (2025) 199} [\href{https://arxiv.org/abs/2503.11395}{{\ttfamily
  2503.11395}}].

\bibitem{Roberge:1986mm}
A.~Roberge and N.~Weiss, \emph{{Gauge Theories With Imaginary Chemical
  Potential and the Phases of {QCD}}},
  \href{https://doi.org/10.1016/0550-3213(86)90582-1}{\emph{Nucl. Phys. B}
  {\bfseries 275} (1986) 734}.

\bibitem{DallaBrida:2026kuo}
M.~Dalla~Brida, R.~H{\"o}llwieser, F.~Knechtli, T.~Korzec, A.~Ramos, S.~Sint
  et~al., \emph{{High-precision calculation of the quark{\textendash}gluon
  coupling from lattice QCD}},
  \href{https://doi.org/10.1038/s41586-026-10339-4}{\emph{Nature} {\bfseries
  652} (2026) 328}.

\bibitem{DallaBrida:2021ddx}
M.~Dalla~Brida, L.~Giusti, T.~Harris, D.~Laudicina and M.~Pepe,
  \emph{{Non-perturbative thermal QCD at all temperatures: the case of mesonic
  screening masses}},
  \href{https://doi.org/10.1007/JHEP04(2022)034}{\emph{JHEP} {\bfseries 04}
  (2022) 034} [\href{https://arxiv.org/abs/2112.05427}{{\ttfamily
  2112.05427}}].

\bibitem{Bresciani:2025vxw}
M.~Bresciani, M.~Dalla~Brida, L.~Giusti and M.~Pepe, \emph{{QCD Equation of
  State with Nf=3 Flavors up to the Electroweak Scale}},
  \href{https://doi.org/10.1103/PhysRevLett.134.201904}{\emph{Phys. Rev. Lett.}
  {\bfseries 134} (2025) 201904}
  [\href{https://arxiv.org/abs/2501.11603}{{\ttfamily 2501.11603}}].

\bibitem{Bresciani:2025mcu}
M.~Bresciani, M.~Dalla~Brida, L.~Giusti and M.~Pepe, \emph{{QCD equation of
  state at very high temperature: Computational strategy, simulations, and data
  analysis}}, \href{https://doi.org/10.1103/jl9n-lk9k}{\emph{Phys. Rev. D}
  {\bfseries 113} (2026) 034506}
  [\href{https://arxiv.org/abs/2511.09160}{{\ttfamily 2511.09160}}].

\bibitem{Wolff:2003sm}
{\scshape ALPHA} collaboration, \emph{{Monte Carlo errors with less errors}},
  \href{https://doi.org/10.1016/S0010-4655(03)00467-3}{\emph{Comput. Phys.
  Commun.} {\bfseries 156} (2004) 143}
  [\href{https://arxiv.org/abs/hep-lat/0306017}{{\ttfamily hep-lat/0306017}}].

\bibitem{Ramos:2018vgu}
A.~Ramos, \emph{{Automatic differentiation for error analysis of Monte Carlo
  data}}, \href{https://doi.org/10.1016/j.cpc.2018.12.020}{\emph{Comput. Phys.
  Commun.} {\bfseries 238} (2019) 19}
  [\href{https://arxiv.org/abs/1809.01289}{{\ttfamily 1809.01289}}].

\bibitem{Joswig:2022qfe}
F.~Joswig, S.~Kuberski, J.T.~Kuhlmann and J.~Neuendorf, \emph{{pyerrors: A
  python framework for error analysis of Monte Carlo data}},
  \href{https://doi.org/10.1016/j.cpc.2023.108750}{\emph{Comput. Phys. Commun.}
  {\bfseries 288} (2023) 108750}
  [\href{https://arxiv.org/abs/2209.14371}{{\ttfamily 2209.14371}}].

\bibitem{Laine:2009dh}
M.~Laine and M.~Vepsalainen, \emph{{On the smallest screening masses in hot
  QCD}}, \href{https://doi.org/10.1088/1126-6708/2009/09/023}{\emph{JHEP}
  {\bfseries 09} (2009) 023} [\href{https://arxiv.org/abs/0906.4450}{{\ttfamily
  0906.4450}}].

\bibitem{Giusti:2018cmp}
L.~Giusti and M.~L\"uscher, \emph{{Topological susceptibility at $T>T_{\rm c}$
  from master-field simulations of the SU(3) gauge theory}},
  \href{https://doi.org/10.1140/epjc/s10052-019-6706-7}{\emph{Eur. Phys. J. C}
  {\bfseries 79} (2019) 207}
  [\href{https://arxiv.org/abs/1812.02062}{{\ttfamily 1812.02062}}].

\bibitem{Bonati:2015vqz}
C.~Bonati, M.~D'Elia, M.~Mariti, G.~Martinelli, M.~Mesiti, F.~Negro et~al.,
  \emph{{Axion phenomenology and $\theta$-dependence from $N_f = 2+1$ lattice
  QCD}}, \href{https://doi.org/10.1007/JHEP03(2016)155}{\emph{JHEP} {\bfseries
  03} (2016) 155} [\href{https://arxiv.org/abs/1512.06746}{{\ttfamily
  1512.06746}}].

\bibitem{Borsanyi:2016ksw}
S.~Borsanyi et~al., \emph{{Calculation of the axion mass based on
  high-temperature lattice quantum chromodynamics}},
  \href{https://doi.org/10.1038/nature20115}{\emph{Nature} {\bfseries 539}
  (2016) 69} [\href{https://arxiv.org/abs/1606.07494}{{\ttfamily 1606.07494}}].

\bibitem{Caracciolo:1988hc}
S.~Caracciolo, G.~Curci, P.~Menotti and A.~Pelissetto, \emph{{The Energy
  Momentum Tensor on the Lattice: The Scalar Case}},
  \href{https://doi.org/10.1016/0550-3213(88)90332-X}{\emph{Nucl. Phys.}
  {\bfseries B309} (1988) 612}.

\bibitem{Caracciolo:1991cp}
S.~Caracciolo, P.~Menotti and A.~Pelissetto, \emph{{One loop analytic
  computation of the energy momentum tensor for lattice gauge theories}},
  \href{https://doi.org/10.1016/0550-3213(92)90339-D}{\emph{Nucl. Phys.}
  {\bfseries B375} (1992) 195}.

\bibitem{Caracciolo:1991vc}
S.~Caracciolo, P.~Menotti and A.~Pelissetto, \emph{{Analytic determination at
  one loop of the energy momentum tensor for lattice QCD}},
  \href{https://doi.org/10.1016/0370-2693(91)91632-6}{\emph{Phys. Lett.}
  {\bfseries B260} (1991) 401}.

\bibitem{Bruno:2022mfy}
M.~Bruno and R.~Sommer, \emph{{On fits to correlated and auto-correlated
  data}}, \href{https://doi.org/10.1016/j.cpc.2022.108643}{\emph{Comput. Phys.
  Commun.} {\bfseries 285} (2023) 108643}
  [\href{https://arxiv.org/abs/2209.14188}{{\ttfamily 2209.14188}}].

\bibitem{Giusti:2016iqr}
L.~Giusti and M.~Pepe, \emph{{Equation of state of the SU(3) Yang–Mills
  theory: A precise determination from a moving frame}},
  \href{https://doi.org/10.1016/j.physletb.2017.04.001}{\emph{Phys. Lett.}
  {\bfseries B769} (2017) 385}
  [\href{https://arxiv.org/abs/1612.00265}{{\ttfamily 1612.00265}}].

\bibitem{openQCD}
M.~L{\"u}scher, \emph{{openQCD: Simulation programs for lattice QCD}},
  \href{https://arxiv.org/abs/{https://luscher.web.cern.ch/luscher/openQCD/}}{{\ttfamily
  {https://luscher.web.cern.ch/luscher/openQCD/}}}.

\bibitem{Sheikholeslami:1985ij}
B.~Sheikholeslami and R.~Wohlert, \emph{{Improved Continuum Limit Lattice
  Action for QCD with Wilson Fermions}},
  \href{https://doi.org/10.1016/0550-3213(85)90002-1}{\emph{Nucl. Phys.}
  {\bfseries B259} (1985) 572}.

\bibitem{Luscher:1996sc}
M.~L{\"u}scher, S.~Sint, R.~Sommer and P.~Weisz, \emph{{Chiral symmetry and
  O(a) improvement in lattice QCD}},
  \href{https://doi.org/10.1016/0550-3213(96)00378-1}{\emph{Nucl. Phys.}
  {\bfseries B478} (1996) 365}
  [\href{https://arxiv.org/abs/hep-lat/9605038}{{\ttfamily hep-lat/9605038}}].

\bibitem{Yamada:2004ja}
{\scshape JLQCD, CP-PACS} collaboration, \emph{{Non-perturbative
  O(a)-improvement of Wilson quark action in three-flavor QCD with plaquette
  gauge action}}, \href{https://doi.org/10.1103/PhysRevD.71.054505}{\emph{Phys.
  Rev.} {\bfseries D71} (2005) 054505}
  [\href{https://arxiv.org/abs/hep-lat/0406028}{{\ttfamily hep-lat/0406028}}].

\bibitem{Rabinowitz1980TheED}
P.~Rabinowitz, \emph{The exact degree of precision of generalized gauss-kronrod
  integration rules}, {\emph{Mathematics of Computation} {\bfseries 35} (1980)
  1275}.

\bibitem{Burgio:1996ji}
G.~Burgio, S.~Caracciolo and A.~Pelissetto, \emph{{Algebraic algorithm for the
  computation of one loop Feynman diagrams in lattice QCD with Wilson
  fermions}}, \href{https://doi.org/10.1016/0550-3213(96)00428-2}{\emph{Nucl.
  Phys.} {\bfseries B478} (1996) 687}
  [\href{https://arxiv.org/abs/hep-lat/9607010}{{\ttfamily hep-lat/9607010}}].

\bibitem{Capitani:1994qn}
S.~Capitani and G.~Rossi, \emph{{Deep inelastic scattering in improved lattice
  QCD. 1. The First moment of structure functions}},
  \href{https://doi.org/10.1016/0550-3213(94)00428-H}{\emph{Nucl. Phys.}
  {\bfseries B433} (1995) 351}
  [\href{https://arxiv.org/abs/hep-lat/9401014}{{\ttfamily hep-lat/9401014}}].

\bibitem{Capitani:2002mp}
S.~Capitani, \emph{{Lattice perturbation theory}},
  \href{https://doi.org/10.1016/S0370-1573(03)00211-4}{\emph{Phys. Rept.}
  {\bfseries 382} (2003) 113}
  [\href{https://arxiv.org/abs/hep-lat/0211036}{{\ttfamily hep-lat/0211036}}].

\bibitem{Yang:2016xsb}
Y.-B.~Yang, M.~Glatzmaier, K.-F.~Liu and Y.~Zhao, \emph{{The 1-loop correction
  of the QCD energy momentum tensor with the overlap fermion and HYP smeared
  Iwasaki gluon}},  \href{https://arxiv.org/abs/1612.02855}{{\ttfamily
  1612.02855}}.

\bibitem{Cheng:1998}
S.H.~Cheng and N.J.~Higham, \emph{{A Modified Cholesky Algorithm Based on a
  Symmetric Indefinite Factorization}},
  \href{https://doi.org/10.1137/S0895479896302898}{\emph{SIAM Journal on Matrix
  Analysis and Applications} {\bfseries 19} (1998) 1097}
  [\href{https://arxiv.org/abs/https://doi.org/10.1137/S0895479896302898}{{\ttfamily
  https://doi.org/10.1137/S0895479896302898}}].

\end{thebibliography}\endgroup

\end{document}